\documentclass[longauth]{aa}  

\usepackage{graphicx}
\usepackage{txfonts}
\usepackage{gensymb}
\usepackage[colorlinks=true,linkcolor=blue,citecolor=blue,urlcolor=blue]{hyperref}
\newcommand{\prot}{$P_{\text{rot}}$}
\newcommand{\mps}{m s$^{-1}$}

\begin{document}

   \title{A hot terrestrial planet orbiting the bright M dwarf L~168-9 unveiled by \textit{TESS}
     \thanks{Partially based on observations made with the HARPS instrument 
       on the ESO 3.6 m telescope under the program 
       IDs 198.C-0838(A), 0101.C-0510(C), and 1102.C-0339(A) at Cerro La Silla (Chile).
       }
     \textsuperscript{,}\thanks{Data (Tables XXX) are only 
       available at the CDS via anonymous ftp to 
       cdsarc.u-strasbg.fr (130.79.128.5)\newline
       or via\newline
       http://cdsarc.u-strasbg.fr/viz-bin/qcat?J/A+A/XXX/XXX
     }
   \textsuperscript{,}\thanks{This paper includes data gathered with the 6.5 meter Magellan Telescopes located at Las Campanas Observatory, Chile.}
   }

   \authorrunning{Astudillo-Defru et al.}
   \titlerunning{L~168-9~b: a nearby hot terrestrial world}

   \author{N. Astudillo-Defru\inst{\ref{conce1}} \and
     R. Cloutier\inst{\ref{cfa},\ref{toronto1}} \and
     S. X. Wang\inst{\ref{carnegieobs}} \and
     J. Teske\inst{\ref{carnegieobs},\ref{hubblefellow}} \and
     R. Brahm\inst{\ref{puc},\ref{ipuc},\ref{mas}} \and
     C. Hellier\inst{\ref{Keele}} \and
     G. Ricker\inst{\ref{mit1}} \and
     R. Vanderspek\inst{\ref{mit1}} \and
     D. Latham\inst{\ref{cfa}} \and
     S. Seager\inst{\ref{mit2}} \and
     J.~N.~Winn\inst{\ref{princeton}} \and
     J. M. Jenkins\inst{\ref{nasa}} \and
     K.~A.~Collins\inst{\ref{cfa}} \and
     K. G.~Stassun\inst{\ref{Vanderbilt}} \and
     C. Ziegler\inst{\ref{dunlap}} \and
     J. M. Almenara\inst{\ref{grenoble}} \and
     D. R. Anderson\inst{\ref{Keele},\ref{warwick}} \and
     E. Artigau\inst{\ref{montreal}} \and
     X. Bonfils\inst{\ref{grenoble}} \and
     F. Bouchy\inst{\ref{geneva}} \and
     C. Brice\~{n}o\inst{\ref{ctio}} \and
     R. P. Butler\inst{\ref{carnegiedtm}} \and
     D. Charbonneau\inst{\ref{cfa}} \and
     D.~M. Conti\inst{\ref{aavso}} \and 
     J. Crane\inst{\ref{carnegieobs}} \and
     I.~J~.M. Crossfield\inst{\ref{mit4},\ref{kansas}} \and 
     M. Davies\inst{\ref{nasa}} \and
     X. Delfosse\inst{\ref{grenoble}} \and
     R. F. D\'iaz\inst{\ref{uba},\ref{iafe}} \and
     R. Doyon\inst{\ref{montreal}} \and
     D. Dragomir\inst{\ref{Albuquerque}} \and
     J.~D. Eastman\inst{\ref{cfa}} \and
     N. Espinoza\inst{\ref{stsi}} \and
     Z. Essack\inst{\ref{mit2}} \and
     F. Feng\inst{\ref{carnegiedtm}} \and
     P. Figueira\inst{\ref{eso}, \ref{caup}} \and
     T. Forveille\inst{\ref{grenoble}} \and
     T. Gan\inst{\ref{tsinghua}} \and 
     A. Glidden\inst{\ref{mit1}} \and
     N. Guerrero\inst{\ref{mit1}} \and
     R. Hart\inst{\ref{cesrc}} \and
     Th. Henning\inst{\ref{mpia}} \and
     E.~P. Horch\inst{\ref{scs}} \and 
     G. Isopi\inst{\ref{ccao}} \and 
     J.~S. Jenkins\inst{\ref{uchile}} \and 
     A. Jord\'an\inst{\ref{uai},\ref{mas}} \and
     J.~F. Kielkopf\inst{\ref{louisville}} \and
     N. Law\inst{\ref{ncarolina}} \and
     C. Lovis\inst{\ref{geneva}} \and
     F. Mallia\inst{\ref{ccao}} \and
     A.~W. Mann\inst{\ref{ncarolina}} \and
     J.~R. de Medeiros\inst{\ref{riogrande}} \and
     C. Melo\inst{\ref{eso}} \and
     R. E. Mennickent\inst{\ref{conce2}} \and
     L. Mignon\inst{\ref{grenoble}} \and
     F. Murgas\inst{\ref{grenoble}} \and
     D.~A. Nusdeo\inst{\ref{gsu}} \and 
     F. Pepe\inst{\ref{geneva}} \and
     H.~M. Relles\inst{\ref{cfa}} \and 
     M. Rose\inst{\ref{nasa}} \and 
     N. C. Santos\inst{\ref{caup}, \ref{porto}} \and
     D. S\'egransan\inst{\ref{geneva}} \and
     S. Shectman\inst{\ref{carnegieobs}} \and
     A. Shporer\inst{\ref{mit1}} \and 
     J. C. Smith\inst{\ref{seti},\ref{nasa}}
     P. Torres\inst{\ref{puc}} \and
     S. Udry\inst{\ref{geneva}} \and
     J. Villasenor\inst{\ref{mit3}} \and
     J~.G. Winters\inst{\ref{cfa}} \and 
     G. Zhou\inst{\ref{cfa}}
   }

   \institute{Departamento de Matem\'atica y F\'isica Aplicadas, Universidad Cat\'olica de la Sant\'isima Concepci\'on, Alonso de Rivera 2850, Concepci\'on, Chile \label{conce1} \\E-mail: nastudillo@ucsc.cl
    \and Center for Astrophysics | Harvard \& Smithsonian , 60 Garden Street, Cambridge, MA 02138, USA \label{cfa}
    \and Dept. of Astronomy \& Astrophysics, University of Toronto, 50 St. George Street, Toronto, ON, M5S 3H4, Canada \label{toronto1}
    \and Observatories of the Carnegie Institution for Science, 813 Santa Barbara Street, Pasadena, CA 91101 \label{carnegieobs}
    \and NASA Hubble Fellow \label{hubblefellow}
    \and Center of Astro-Engineering UC, Pontificia Universidad Cat\'olica de Chile, Av. Vicu\~{n}a Mackenna 4860, 7820436 Macul, Santiago, Chile \label{puc}
    \and Instituto de Astrof\'isica, Facultad de F\'isica, Pontificia Universidad Cat\'olica de Chile, Av. Vicuña Mackenna 4860, 7820436 Macul, Santiago, Chile \label{ipuc}
    \and Millennium Institute of Astrophysics, Chile \label{mas}
    \and Astrophysics Group, Keele University, Staffordshire, ST5 5BG, UK \label{Keele}
    \and Kavli Institute for Astrophysics and Space Research, Massachusetts Institute of Technology, Cambridge, MA 02139, USA \label{mit1}
    \and Department of Earth, Atmospheric, and Planetary Sciences, Massachusetts Institute of Technology,  Cambridge,  MA 02139, USA \label{mit2}
    \and Department of Astrophysical Sciences, Princeton University, Princeton, NJ 08544, USA \label{princeton}
    \and NASA  Ames  Research  Center,  Moffett  Field,  CA  94035, USA \label{nasa}
    \and Department of Physics and Astronomy, Vanderbilt University, Nashville, TN 37235, USA \label{Vanderbilt} 
    \and Dunlap Institute for Astronomy and Astrophysics, University of Toronto, Ontario M5S 3H4, Canada \label{dunlap}
     \and Univ. Grenoble Alpes, CNRS, IPAG, F-38000 Grenoble, France \label{grenoble}
     \and Department of Physics, University of Warwick, Gibbet Hill Road, Coventry CV4 7AL, UK \label{warwick}
     \and Institut de Recherche sur les Exoplan\`etes, D\'epartement de Physique, Universit\'e de Montr\'eal, Montr\'eal QC, H3C 3J7, Canada \label{montreal}
     \and Observatoire de Gen\`eve, Universit\'e de Gen\`eve, 51 ch. des Maillettes, 1290 Sauverny, Switzerland \label{geneva}
     \and Cerro Tololo Inter-American Observatory, Casilla 603, La Serena, Chile \label{ctio}
    \and Department of Terrestrial Magnetism, Carnegie Institution for Science, 5241 Broad Branch Road NW, Washington DC 20015, USA \label{carnegiedtm}
    \and American Association of Variable Star Observers, 49 Bay State Road, Cambridge, MA 02138, USA \label{aavso}
     \and Department of Physics, and Kavli Institute for Astrophysics and Space Research, Massachusetts Institute of Technology, Cambridge, MA, USA \label{mit4}
     \and Physics \& Astronomy Department, University of Kansas, Lawrence, KS, 66044, USA \label{kansas}
     \and Universidad de Buenos Aires, Facultad de Ciencias Exactas y Naturales. Buenos Aires, Argentina\label{uba}
    \and CONICET - Universidad de Buenos Aires. Instituto de Astronom\'ia y F\'isica del Espacio (IAFE). Buenos Aires, Argentina \label{iafe}
    \and Department of Physics and Astronomy, University of New Mexico, Albuquerque, NM, USA \label{Albuquerque}
    \and Space Telescope Science Institute, 3700 San Martin Drive, Baltimore, MD 21218 \label{stsi}
    \and European Southern Observatory, Alonso de C\'ordova 3107, Vitacura, Regi\'on Metropolitana, Chile \label{eso}
    \and Universidad de Concepci\'on, Departamento de Astronom\'ia, Casilla 160-C, Concepci\'on, Chile \label{conce2}
    \and Physics Department and Tsinghua Centre for Astrophysics, Tsinghua University, Beijing 100084, China \label{tsinghua}
    \and Centre for Astrophysics, University of Southern Queensland, Toowoomba, QLD, 4350, Australia \label{cesrc}
    \and Max-Planck-Institut fur Astronomie, K\"unigstuhl 17, D-69117 Heidelberg, Germany \label{mpia}
    \and Department of Physics, Southern Connecticut State University, 501 Crescent Street, New Haven, CT 06515, USA \label{scs}
    \and Campo Catino Astronomical Observatory, Regione Lazio, Guarcino (FR), 03010 Italy \label{ccao}
    \and Departamento de Astronom\'ia, Universidad de Chile, Camino El Observatorio 1515, Las Condes, Santiago, Chile \label{uchile}
    \and Facultad de Ingeniería y Ciencias, Universidad Adolfo Ib\'a\~nez, Av.\ Diagonal las Torres 2640, Pe\~nalol\'en, Santiago, Chile \label{uai}
    \and Department of Physics and Astronomy, University of Louisville, Louisville, KY 40292, USA \label{louisville}
    \and Department of Physics and Astronomy, The University of North Carolina at Chapel Hill, Chapel Hill, NC 27599-3255, USA \label{ncarolina}
    \and Departamento de F\'isica, Universidade Federal do Rio Grande do Norte, 59078-970 Natal, RN, Brazil \label{riogrande}
    \and Georgia State University, Department of Physics \& Astronomy, 25 Park Place NE \#605 \label{gsu}
    %\and Leidos, Inc., Moffett Field, CA 94035, USA \label{leidos}
    \and Instituto de Astrof\'isica e Ci\^encias do Espa\c{c}o, Universidade do Porto, CAUP, Rua das Estrelas, PT4150-762 Porto, Portugal \label{caup}
    \and Departamento de F\'isica e Astronomia, Faculdade de Ci\^encias, Universidade do Porto, Portugal \label{porto}
    \and SETI Institute, Moffett Field, CA 94035, USA \label{seti}
    \and Center for Space Research, MIT, 37-414, Cambridge, MA 02139, UA \label{mit3}
    }

   \date{}

% \abstract{}{}{}{}{} 
% 5 {} token are mandatory

  % context heading (optional)
  %{}
  % aims heading (mandatory)
   \abstract{We report the detection of a transiting super-Earth-sized planet
   (R=1.39$\pm$0.09 R$_\oplus$) in a 1.4-day orbit around L~168-9 (TOI-134),
   a bright M1V dwarf (V=11, K=7.1) located at 25.15$\pm$0.02 $pc$. The host star was
   observed in the first sector of the Transiting Exoplanet Survey Satellite (TESS) mission
   and, for confirmation and planet mass measurement, was followed up with ground-based
   photometry, seeing-limited and high-resolution imaging, and precise radial velocity (PRV)
   observations using the HARPS and Magellan/PFS spectrographs. Combining the TESS data 
   and PRV observations, we find the mass of L~168-9~b to be 4.60$\pm$0.56 M$_\oplus$, 
   and thus the bulk density to be $1.74^{+0.44}_{-0.33}$ times larger than that of the Earth.
   The orbital eccentricity is smaller than 0.21 (95\% confidence).  This planet is a Level
   One Candidate for the TESS Mission’s scientific objective -- to measure the masses of 
   50 small planets -- and is one of the most observationally accessible terrestrial planets for
   future atmospheric characterization.}


   \maketitle
%-------------------------------------------------------------------
\section{Introduction}

The best planets for detailed characterization are transiting planets, first and foremost
because they allow for the possibility of unambiguous mass measurement 
\citep[e.g., HD209458b][]{Charbonneau2000, Mazeh2000, Henry2000}.
From the Doppler effect we can determine the minimum
mass of the planet ($m \sin i$), and from the transit light curve we can determine
the planetary radius and the orbital inclination ($i$), thus yielding a measurement of 
the planet's mass. Moreover, from this combination we can calculate the planet's mean
density, and shed light on its internal structure by comparison with models containing
different amount of iron, silicates, water, hydrogen, and helium. Furthermore, transiting
planets are unique because of the  feasibility to characterize the upper atmosphere by
spectroscopy during transits and occultations of a significant number of planets.  
The forthcoming \textit{James Webb Space Telescope} \cite[JWST,][]{Gardner2006} and the
\textit{Extremely Large Telescope} \citep[ELT,][]{deZeeuw2014} will have unprecedented
capabilities for detailed studies of the atmospheres of terrestrial planets, and the
interpretation of the results will require an accurate mass measurement (e.g., \citealt{Batalha:2019}).

The Transiting Exoplanet Survey Satellite (TESS) \citep{Ricker2015} started scientific
operations in July 2018, aiming to detect transiting planets around bright and nearby stars
-- bright enough for Doppler mass measurement to be feasible.
For this task, TESS surveys about 85\% of the sky during the Prime Mission.
The survey covering the southern ecliptic hemisphere is now complete,
and the northern survey is underway.  Each hemisphere is divided into
13 rectangular sectors of $96\degree \times 24\degree$ each. Each sector is
continuously observed for an interval of 27-days, with a cadence of 2 minutes for several hundred
thousand pre-selected stars deemed best suited for planet searching.
Additionally, during the TESS Prime Mission, the Full Frame
Images -- the full set of all science and collateral pixels across all CCDs of a given 
camera -- are available with a cadence of 30 minutes.  M dwarfs are of special interest 
because the transit and radial-velocity signals of a given type of planet are larger for
these low-mass stars than they are for Sun-like stars. In addition, compared to hotter stars
M dwarfs present better conditions for the detection of planets orbiting the circumstellar
habitable zone:
less time consuming, larger Doppler signals, and an increased transit probability.
\citet{Sullivan2015} anticipated that from 556 small ($<$2R$_\oplus$) transiting planets
discovered by TESS, 23\% of them will be detected orbiting bright (K$_S<$9) stars, and
that 75\% of small planets will be found around M dwarfs.

This paper reports the discovery of a small planet orbiting the star L~168-9 (TOI-134),
based on TESS data. The host star is a bright M dwarf. An intense precise radial-velocity
campaign with HARPS and the Planet Finder Spectrograph (PFS) revealed the terrestrial 
nature of the newly detected world. This work is presented as follows:
Section~\ref{sec:L168-9} describes the host star properties. Sections~\ref{sec:phot}
and~\ref{sec:rv} describe the photometric and radial-velocity observations.
Section~\ref{sec:analysis} 
presents an analysis of all the data, including the study of
stellar activity.  Finally, Section~\ref{sec:conclusion} places L~168-9~b within the larger
context of the sample of detected planets.

%--------------------------------------------------------------------
\section{L~168-9}
\label{sec:L168-9}

\begin{table}[t]
  \caption{L~168-9 (TIC~234994474) properties. Superscripts indicate the reference.}
  \label{tab:stellarprop}
  \centering
  \small
  \begin{tabular}{p{0.2\linewidth}l c c c}  
    \hline\noalign{\smallskip}
    Parameter & Units & Value & Reference \\
    \noalign{\smallskip}
    \hline\noalign{\medskip}
    R.A. & [J2000] & $23^h 20^m 07.52^s$  & Gaia2018\\
    Decl. & [J2000] & $-60\degree 03^\prime 54.64^{\prime\prime}$ &  Gaia2018\\
    Spectral type    &  &  M1V & Ga2014\\
    B & [mag] & 12.45$\pm$0.19 & Ho2000\\
    V & [mag] & 11.02$\pm$0.06 & Ho2000\\
    B$_A$ & [mag] & 12.460$\pm$0.025 & He2016\\
    V$_A$ & [mag] & 11.005$\pm$0.018 & He2016\\
    g$_A$ & [mag] & 11.752$\pm$0.032 & He2016\\
    r$_A$ & [mag] & 10.416$\pm$0.028 & He2016\\
    i$_A$ & [mag] & 9.675 & He2016\\
    W$_1$ & [mag] & 6.928$\pm$0.060 & Cu2013\\
    W$_2$ & [mag] & 6.984$\pm$0.020 & Cu2013\\
    W$_3$ & [mag] & 6.906$\pm$0.016 & Cu2013\\
    W$_4$ & [mag] & 6.897$\pm$0.074 & Cu2013\\
    T && 9.2298$\pm$0.0073 & St2018\\
    J & &7.941$\pm$0.019 & Cu2003\\         
    H & &7.320$\pm$0.053 & Cu2003\\         
    K$_s$ & &7.082$\pm$0.031 & Cu2003\\
    B$_p$ & &11.2811$\pm$0.0016 & Gaia2018\\
    G & & 10.2316$\pm$0.0008 & Gaia2018\\
    R$_p$ & &9.2523$\pm$0.0011 & Gaia2018\\
    $\pi$ &  [mas]   &  39.762$\pm$0.038 & Gaia2018\\
    Distance & [pc] & 25.150 $\pm$ 0.024 & Gaia2018\\
    $\mu_\alpha$ & [mas/yr] & -319.96$\pm$0.10 & Gaia2018\\
    $\mu_\delta$ & [mas/yr] &  -127.78$\pm$0.12 & Gaia2018\\
    $dv_r/dt$ &[m/s/yr] & 0.06865$\pm$0.00011 & this work\\
    M$_s$ & [M$_\odot$] & 0.62$\pm$0.03 & Ma2019\\
    R$_s$ & [R$_\odot$] & 0.600 $\pm$0.022 & Sect.~\ref{subsec:stellarparam}\\
    T$_{\rm eff}$ & [K] & 3800$\pm$70 & Sect.~\ref{subsec:stellarparam}\\
    L$_s$ & [L$_{\odot}$] & 0.0673$\pm$0.0024 & Sect.~\ref{subsec:stellarparam}\\
    ${log(g)}$ & [g/cm$^{-3}$]  &4.04$\pm$0.49 & Sect.~\ref{subsec:stellarparam}\\
    $[\mathrm{Fe/H}]$ &  & 0.04$\pm$0.17 & Ne2014\\
    ${log(R^\prime_{HK})}$ &  & -4.562$\pm$0.043 & As2017A\\
    P$_{\text{rot}}$ & [days] & $29.8\pm1.3$ & Sect.~\ref{subsec:rotation}\\
    \noalign{\smallskip}\hline
  \end{tabular}

  \begin{list}{}{}
  \item[Reference notes: ] Gaia2018~--~\citet{Gaia2018}; 
  Ga2014~--~\citet{Gaidos2014};
  Ho2000~--~\citet{Tycho-2};
  H22016~--~\citet{apass};
  Cu2013~--~\citet{wise};
  St2018~--~\citet{StassunTIC:2018};
  Cu2003~--~\citet{Cutri2003}; 
  Ma2019~--~\citet{Mann:2019};
  Ne2014~--~\citet{Neves2014};
  As2017a~--~\citet{Astudillo2017a}
  \end{list}
\end{table}

L~168-9, also known as CD-60~8051, HIP~115211, 2MASS~J23200751-6003545, with 
the entry 234994474 of the TESS Input Catalog (TIC) or 134 of the Tess Object of Interest (TOI) list, 
is a red dwarf of spectral type M1V.
It appears in the southern sky, and resides at a distance of 25.150$\pm$0.024 $pc$ from the
Sun \citep{Gaidos2014, Gaia2018}. 
Table~\ref{tab:stellarprop} lists the key parameters of the star, namely 
its position, visual and near-infrared apparent magnitudes, parallax, proper motion, 
secular acceleration, and its essential physical properties.

\subsection{Derived stellar properties}
\label{subsec:stellarparam}

\begin{figure}
\centering
%\sidecaption    
    \includegraphics[clip, trim=3.6cm 2.6cm 3cm 3cm,scale=0.4]{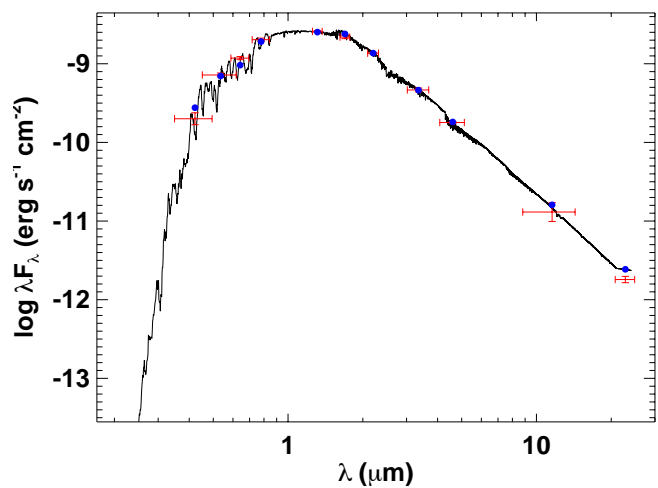}
    \caption{Spectral energy distribution (SEDs) of L~168-9. Red error bars represent the 
observed photometric measurements, where the horizontal bars represent the effective 
width of the passband. Blue circles are the model fluxes from the best-fit NextGen 
atmosphere model (black). 
    \label{fig:sed}}
\end{figure}

We performed an analysis of the broadband spectral energy distribution (SED) 
together with the {\it Gaia\/} DR2 parallax in order to determine an empirical 
measurement of the stellar radius, following the procedures described by
\citet{Stassun:2016,Stassun:2017,Stassun:2018}. We took the $B_T V_T$ 
magnitudes from {\it Tycho-2}, the $BVgri$ magnitudes from APASS, 
the $JHK_S$ magnitudes from {\it 2MASS}, the W1--W4 magnitudes from 
{\it WISE}, and the $G$ magnitude from {\it Gaia}. Together, the 
available photometry spans the full stellar SED over the wavelength 
range 0.35--22~$\mu$m (see Figure~\ref{fig:sed}).

We performed a fit using NextGen stellar atmosphere models \citep{Hauschildt1999},
with the effective temperature ($T_{\rm eff}$) and surface gravity ($\log g$) constrained 
on the ranges reported in the TESS Input Catalog \citep{StassunTIC:2018}, while the metallicity
[Fe/H] was fixed to a typical M-dwarf metallicity of -0.5. 
We fixed the extinction ($A_V$) to be zero, considering proximity of the star, the
degrees-of-freedom of the fit is 10. 
The resulting fit (Figure~\ref{fig:sed}) has a $\chi^2$ of 42.3 ($\chi^2_{red}$=4.2), with 
$T_{\rm eff} = 3800 \pm 70$~K. The relatively high $\chi^2$ is likely due to systematics, 
as the stellar atmosphere model is not perfect. We artificially increased
the observational uncertainty estimates until $\chi^2_{red}=1$ was achieved.
Integrating the (unreddened) model 
SED gives the bolometric flux at Earth of $F_{\rm bol} = 3.41 \pm 0.12 \times 10^{-9}$
erg~s$^{-1}$~cm$^{-2}$. 
Taking the $F_{\rm bol}$ and $T_{\rm eff}$ together with the {\it Gaia\/} DR2 parallax, 
adjusted by $+0.08$~mas to account for the systematic offset reported by 
\citet{StassunTorres:2018}, gives the stellar radius as $R = 0.600 \pm 0.022$~R$_\odot$. 
Finally, estimating the stellar mass from the empirical relations of \citet{Mann:2019} 
gives $M = 0.62 \pm 0.03 M_\odot$. With these values of the mass and radius,
the stellar mean density is $\rho = 4.04 \pm 0.49$ g~cm$^{-3}$. We also tested to fit 
the SED using BT-Settl theoretical grid of stellar model \citep{Allard2014}, where we 
obtained a consistent result.

We searched for infrared (IR) excess in {\it WISE} data using the Virtual Observatory SED
Analyser \citep[VOSA,][]{Bayo:2008}, which could point for the presence of debris disks. 
For that we computed the excess significance parameter
\citep[$\chi_\lambda$,][]{Beichman:2006,Moor:2006}, where $\chi_\lambda \ge 3$ 
represents a robust detection of IR excess. We obtained  $\chi_\lambda = 0.70$ in the
W$_3$ band, ruling out the presence of a debris disk around L~168-9.

%-------------------------------------------------------------------

\section{Observations}

The first hint of a planetary companion orbiting L~168-9 came from analyzing TESS data.
After the Data Validation Report was released to the community, a follow-up campaign started
with several instruments and by different teams to check on whether the transit-like signal
seen by TESS originated from a planet, as opposed to a stellar binary or other source.
The follow-up observations included supplementary time series photometry aiming to detect 
additional transits, seeing-limited and high-resolution imaging to analyze the possibility 
that the signal comes from a star on a nearby sightline, and precise radial-velocity monitoring
to measure the companion's mass.

\subsection{Photometry}
\label{sec:phot}

\subsubsection{TESS}

TESS observed Sector 1 from the 25th of July to the 22nd of August 2018
\footnote{The Sector 1 pointing direction was $RA(J2000):+352.68\degree,\
  Dec(J2000):-64.85\degree,\ Roll:-137.85\degree$.},
a 27.4-day interval that is typical of each sector. L~168-9 is listed in the Cool Dwarf Catalog 
that gathered the known properties of as many dwarf stars as possible with $V-J>2.7$ and effective
temperatures lower than 4\,000 K \citep{Muirhead2018}.
The predicted TESS-band apparent magnitudes are also given in this catalog.
L~168-9 was chosen for 2-min time sampling as part of the TESS Candidate Target List
\citep[CTL,][]{StassunTIC:2018}, consisting of a subset of the TESS Input Catalog (TIC)
identified as high-priority stars in the search for small transiting
planets. Time series observations of L~168-9 were made with CCD 2 of Camera 2.

The TESS Science Processing Operations Center (SPOC) at the NASA Ames Research Center
performed the basic calibration, reduction, and de-trending of the time series, and also
performed the search for transit-like signals \citep{Jenkins2016}. The light curves were
derived by the SPOC pipeline and consist of a time series based on Simple Aperture 
Photometry (SAP), as well as a corrected time series based on Pre-search Data Conditioning
\citep[PDC,][]{Smith2012,Stumpe2014} referred to as PDCSAP (as detailed by Tenenbaum and
Jenkins 2018\footnote{\url{https://archive.stsci.edu/missions/tess/doc/EXP-TESS-ARC-ICD-TM-0014.pdf}}).
This work made use of the PDCSAP time series available on the Mikulski Archive for Space
Telescopes
(MAST\footnote{\url{https://mast.stsci.edu/portal/Mashup/Clients/Mast/Portal.html}}).
The TESS photometry is presented in the upper panel in Figure~\ref{fig:tess_phot}.

A Data Validation Report for L~168-9 was released to the community as part of the MIT TESS
Alerts\footnote{\url{https://tev.mit.edu/toi/alerts/}}. The report includes 
several validation tests to assess the probability that the signal is a false positive:
eclipsing-binary discrimination tests, a statistical bootstrap test, a ghost diagnostic
test, and difference-image centroid offset tests. These are described by
\citet{Twicken2018}. All the tests were passed successfully. The formal false-alarm 
probability of the planet candidate was reported to be 5.85$\times 10^{-37}$.

The TESS time series covers 19 transits of what was originally deemed a planet candidate
(L~168-9\,b or TOI-134.01), and reported on the TESS exoplanet Follow-up Observing Program
(TFOP) website\footnote{\url{https://exofop.ipac.caltech.edu/tess/}}. According to the 
Data Validation Report the orbital period is P~$=1.401461\pm0.000137$ days and
the transit depth is $\Delta F/F_0=566\pm38\ ppm$, which translates into a planetary radius 
of R$_p=1.58\pm0.36$ R$_\oplus$ (Sect.~\ref{subsec:stellarparam} describes how we determined
the stellar radius, which is the same value as the used in the Validation Report). 
The time of mid-transit at an arbitrarily chosen reference epoch is (BJD)
T$_c=2458326.0332\pm0.0015$.

\subsubsection{LCOGT, MKO, and SSO T17}
\label{subsubsec:groundphot}

We acquired ground-based time series photometric follow-up of L~168-9 and the nearby field
stars as part of the TESS Follow-up Observing Program (TFOP) to attempt to rule out nearby
eclipsing binaries (NEBs) in all stars that are bright enough to cause the TESS detection
and that could be blended in the TESS aperture. We used the 
{\tt TESS Transit Finder}\footnote{\url{https://astro.swarthmore.edu/telescope/tess-secure/find\_tess\_transits.cgi}}, 
which is a customized version of the {\tt Tapir} software package \citep{Jensen:2013}, to schedule our
transit observations.

We observed one full transit simultaneously using three 1-meter telescopes at the Las Cumbres
Observatory Global Telescope (LCOGT) \citep{Brown2013} South Africa Astronomical Observatory
node on 21 September 2018 in $i'$-band. The Sinistro detectors consist of 
4K$\times$4K 15-$\mu m$ pixels with an image scale of $0\farcs389$  pixel$^{-1}$, 
resulting in a field-of-view of $26\farcm5 \times 26\farcm5$.
The images were calibrated by the standard LCOGT BANZAI pipeline. 

We observed a full transit from the Mount Kent Observatory (MKO) 0.7-meter telescope near
Toowoomba, Australia on 23 September 2018 in $r'$-band. The Apogee U16 detector consists of
4K$\times$4K 9-$\mu m$ pixels with an image scale of $0\farcs41$  pixel$^{-1}$, resulting in
a field-of-view of $27\arcmin\times27\arcmin$. The images were calibrated using 
{\tt AstroImageJ} ({\tt AIJ}) software package \citep{Collins:2017}.

We observed a full transit from the Siding Spring Observatory, Australia, iTelescope T17
0.43-meter telescope on 27 September 2018 with no filter. The FLI ProLine PL4710 detector
consists of 1K$\times$1K pixels with an image scale of $0\farcs92$  pixel$^{-1}$, 
resulting in a field-of-view of $15\farcm5\times15\farcm5$. The images were calibrated using 
{\tt AstroImageJ}.

We used the {\tt AstroImageJ} to extract differential light curves of L~168-9 and all known
stars within $2\farcm5$ of the target star that are bright enough to have possibly produced the
shallow TESS detection, which includes 11 neighbors brighter than TESS-band = 17.9 mag. 
This allows an extra 0.5 in delta magnitude relative to L~168-9 to account for any inaccuracies
in the TESS band reported magnitudes in the TICv8. The L~168-9 light curves in all five
photometric data sets show no significant detection of the shallow TESS detected event, as
expected from our lower precision ground-based photometry. Considering a combination of all five
photometric data sets, we exclude all 11 known neighbors that are close enough and bright enough
to L~168-9 to have possibly caused the TESS detection as potential sources of the TESS
detection. 

\subsubsection{WASP}

WASP-South, located in Sutherland, South Africa, is the southern station of the 
\textit{Wide Angle Search for Planets} \citep[WASP,][]{Pollacco2006}.  It consists 
of an array of 8 cameras each backed by a 2048x2048 CCD. Observations in 2010 and 2011 
(season A) used 200mm, f/1.8 lenses with a broadband filter spanning $400-700$ nm and a 
plate scale of $13.7\arcsec$/pixel. Then, from 2012 to 2014 (season B), WASP-South used 
85mm, f/1.2 lenses with a Sloan r' filter and a  plate scale of $32\arcsec$/pixel. 
The array rastered a number of fields each clear night at typically 10-min cadence.

L~168-9 was monitored for four consecutive years, from to May 20, 2010 to December 12, 2014;
typically covering 150 days in each year. In one campaign two cameras with overlapping
fields observed the star, giving a total of 27\,300 data points; in another campaign,
L~168-9 was observed by three cameras with overlapping fields, totalling 170\,000 data
points. The photometry has a dispersion of 0.027 $\delta$mag and average uncertainty of 
0.024 $\delta$mag, presenting clear signs of variability, as shown below in
Sec.~\ref{subsec:rotation}.

\subsection{High-resolution Imaging}
\label{sec:image}
The relatively large 21\arcsec pixels of TESS can lead to photometric contamination from 
nearby sources. These must be accounted for to rule out astrophysical false positives, 
such as background eclipsing binaries, and to correct the estimated planetary radius, 
initially derived from the diluted transit in a blended light curve \citep{ziegler18}. 
Without this correction, the interpreted planet radius can be underestimated 
\citep[e.g.,][]{ciardi2015, Teske2018}.

\subsubsection{SOAR}
\label{subsubsec:soar}

\begin{figure}[t]
\centering
\includegraphics[scale=0.5]{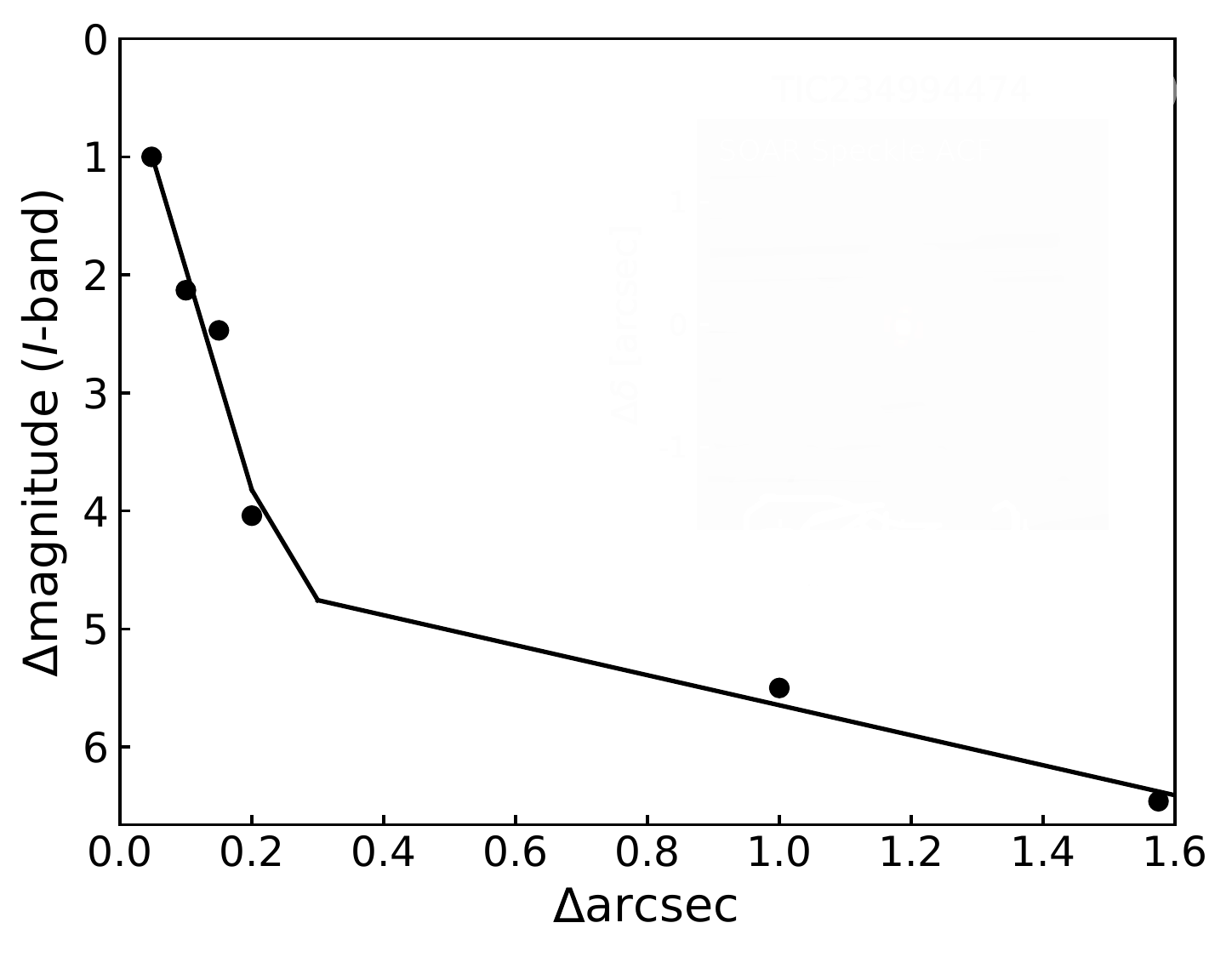}
\caption{SOAR speckle results of L~168-9. Black points represent the I-band contrast 
obtained at a given separation of the star. The solid black line shows the $5 \sigma$
detection limit curve.}
\label{fig:soar}
\end{figure}

 We searched for close companions to L~168-9 with speckle imaging on the 
 4.1-m Southern Astrophysical Research telescope \citep[SOAR,][]{tokovinin18} installed 
 in Cerro Pach\'on, Chile, on 2018 September 25 UT using the I-band 
 ($\lambda_{cen}=824\ nm$, full width at half maximum$=170\ nm$) centered  approximately on 
 the TESS passband. Further details of the TESS SOAR survey are published in \cite{Ziegler2019}.
 
 Figure \ref{fig:soar} shows the 5$\sigma$ detection sensitivity. No nearby stars 
 ($\rho < 1 \farcs 6$) to L~168-9 were detected within the sensitivity
 limits of SOAR.

\subsubsection{Gemini-South}
\label{subsubsec:gemini}

\begin{figure}[t]
\centering
\includegraphics[scale=1.8]{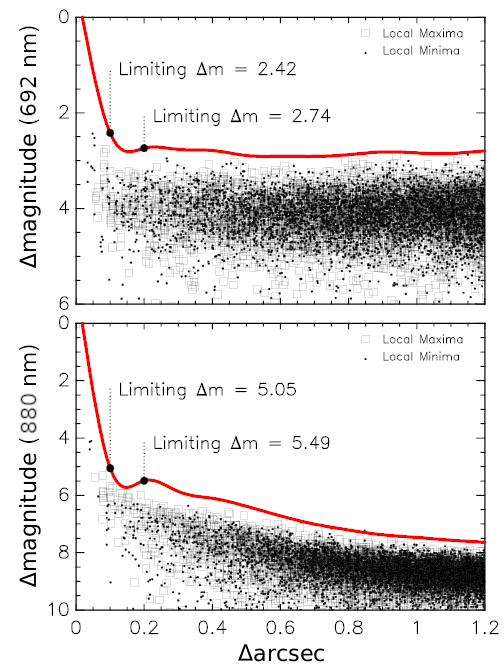}
\caption{DSSI/Gemini-S detection limit curves of L~168-9 in the 692 nm (top) and 880 nm 
(bottom) filters. Squares and points on left panel represent local maxima and local minima, respectively.
}
\label{fig:speckle}
\end{figure}

 Observations of L~168-9 were conducted with the Differential Speckle Survey Instrument 
 (DSSI; \citealt{Horch2009}) on Gemini South, Chile, on UT 28 October 2018 under program 
 GS-2018B-LP-101 (PI: I. Crossfield). The usual 692 nm and 880 nm filters on DSSI were used, 
 and three sequences of 60 ms/frame$\times$1000 frames were taken. The total time on target, 
 including readout overhead, was six minutes. \cite{Howell2011}, \cite{Horch2011a}, 
 and \cite{Horch2012} detail the speckle observing and data reduction procedures. 
 The detection limit curves are shown in Figure~\ref{fig:speckle}. 
 While the 692 nm image was taken at too low of gain, leading to a shallow detection limit, 
 the 880 nm detection limit curve ($b$) in Figure~\ref{fig:speckle} indicates that 
 L~168-9 lacks any companions of $\Delta m \sim$5.0 mag beyond 0.1\arcsec 
 and any companions of $\Delta m \sim$5.5 mag beyond 0.2\arcsec. Gemini-South and 
 SOAR result (Sect.~\ref{subsubsec:soar}) mean the transit signal is likely to be 
 associated with L~168-9.

Thus, we conclude that there is no significant contamination of the TESS photometric
aperture that would bias the determination of the planet radius. 
The Sinistro data (Sect.~\ref{subsubsec:groundphot}) rule out surrounding stars as 
potential sources of the TESS detection, so we can assume the planet orbits L~168-9. 
Given the limits placed on nearby companions by the Gemini-South data, if there were 
a close companion it would have to be fainter by $\sim 5$ magnitudes than the primary star,
meaning the planet radius correction factor would be at most
\begin{equation}
    R_{p,corr} = R_{p}\times \sqrt{1 + 10^{-0.4\Delta m}} = R_{p}\times1.005.
\end{equation}

\noindent This is smaller than the derived planet radius uncertainty (Table~\ref{tab:results}).

\subsection{Radial velocity}
\label{sec:rv}

\subsubsection{HARPS}
The \textit{High Accuracy Radial velocity Planet Searcher} \citep[HARPS][]{Mayor2003} 
is an echelle spectrograph mounted on the 3.6m telescope at La Silla Observatory, Chile.
The light is spread over two CCDs (pixel size 15 $\mu m$) by a science fiber and a
calibration fiber. The calibration fiber can be illuminated with the calibration
lamp for the best radial velocity precision, or it can be placed on sky for moderate
precision. HARPS is stabilized in pressure and temperature, and has a resolving
power of 115{,}000.  The achievable precision in radial velocity of better than 1~$m/s$.

We began monitoring L~168-9 with HARPS on September 29, 2018, soon after the
TESS alert. We elected not to use the simultaneous wavelength calibration 
(i.e.\ the on-sky calibration fiber) to ensure that the bluer spectral regions 
would not be contaminated by the calibration lamp, that provides a much stronger flux 
for any instrumental setup.. The exposure time was set to 900 s for ESO programs 
198.C-0838 and 1102.C-0339, and to 1{,}200 s for ESO program 0101.C-0510, translating 
in a median signal-to-noise ratio per spectral pixel of 51 and 70 at 650 nm, respectively. 
A single spectrum on October 1, 2018, had an exposure time of 609 s for an unknown reason. 
A total of 47 HARPS spectra were collected, ending with observations on December 19, 2018.

HARPS spectra were acquired in roughly three packs of data separated in time by about 
40 days. This sampling is reflected in the window function presented in
Figure~\ref{fig:GLSP}. Two archival spectra of L~168-9 are available at the ESO database. 
However, a radial velocity offset was introduced on May 2015 because the vacuum vessel was
opened during a fiber upgrade \citep{LoCurto2015}. As this offset is not yet well
characterized for M dwarfs, we decided to disregard those two points (from July 2008 and
June 2009) in our subsequent analysis.

The HARPS Data Reduction Software \citep{LovisPepe2007} computes radial velocities
by a cross-correlation function technique \citep[e.g.,]{Baranne:1996}. Nevertheless 
we derived radial velocities by a different approach to exploit as much as possible the
Doppler information of spectra \citep[e.g.,][]{Anglada2012}.
We performed a maximum likelihood analysis between a stellar template and each individual
spectrum following the procedure presented in \citet{Astudillo-Defru2017b}. The stellar
template corresponds to a true stellar spectrum of the star, enhanced in signal-to-noise. 
It was made from the median of all the spectra, after shifting them into a common
barycentric frame. The resulting template is Doppler shifted by a range of trial radial
velocities to construct the maximum likelihood function, from which we derived the HARPS
radial velocity used in the subsequent analysis. The obtained radial velocities -- listed 
in Table~\ref{tab:rvHARPS} -- present a dispersion of 4.01 $m/s$ and a median photon 
uncertainty of 1.71 $m/s$. Figure~\ref{fig:phasedRV} shows the radial velocities 
folded to the orbital period.

\subsubsection{PFS}

% Intro to PFS instrument
The Planet Finder Spectrograph is an iodine-calibrated, environmentally controlled high resolution 
PRV spectrograph \citep{Crane:2006,Crane:2008,Crane:2010}. Since first light 
in October 2009, PFS has been running a long-term survey program to search for 
planets around nearby stars (e.g., ~\citealt{Teske:2016}). In January 2018, 
PFS was upgraded with a new large format CCD with $9\mu m$ pixels and switched 
to a narrower slit for its regular operation mode to boost the resolution from 
80{,}000 to 130{,}000. The PFS spectra are reduced and analyzed with 
a custom IDL pipeline that is capable of delivering RVs with $<$1~m/s precision 
\citep{Butler:1996}.

% The basic and essential description of the observations
We followed up L~168-9 with PFS on the 6.5~m Magellan II Clay telescope 
at Las Campanas Observatory in Chile from October 13--26, and then 
on December 16 and 21 in 2018. Observations were conducted on 15 nights,
with multiple exposures per night. There were a total of 76 exposures
of 20 minutes each. We typically took 2--6 exposures per night over a range of timescales,
to increase the total SNR per epoch and also to average 
out the stellar and instrumental jitter. Each exposure had a typical 
SNR of 28 per pixel near the peak of the blaze function, or 56 per resolution element. 
The radial velocity dispersion is 4.61 m/s and the median RV uncertainty per exposure is 
about 1.8~m/s. Five consecutive 20-minute iodine-free exposures were obtained to allow for 
the construction of a stellar spectral template in order to extract the RVs. These were
bracketed with spectra of rapidly rotating B stars taken through the iodine cell, for
reconstruction of the spectral line spread function and wavelength calibration for the 
template observations.

% Intro to MTS
The PFS observations of L~168-9 presented here are part of the Magellan 
TESS Survey (MTS) that will follow up $\sim$30 super-Earths and sub-Neptunes 
discovered by TESS in the next three years using PFS (Teske et al.~in prep.). 
The goal of MTS is to conduct a statistically robust survey to understand the 
formation and evolution of super-Earths and sub-Neptunes. The observation 
schedules of all MTS targets, including L~168-9, can be found on the 
ExoFOP-TESS website.\footnote{\url{https://exofop.ipac.caltech.edu/tess/}}

\section{Analysis}
\label{sec:analysis}

\begin{figure*}
    \sidecaption
    %\centering
    \includegraphics[scale=0.5]{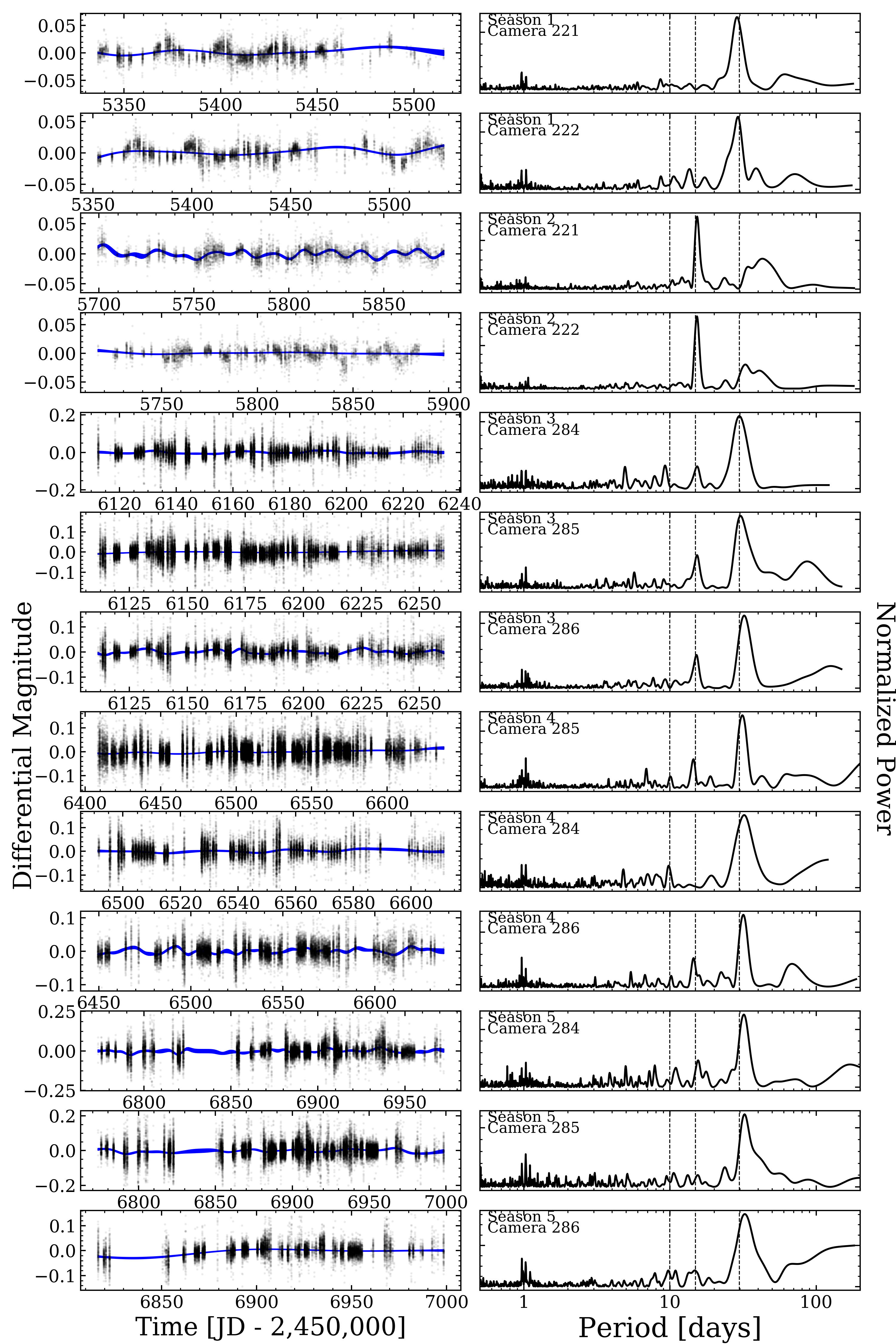}
    \caption{The WASP photometry. \emph{Left column}: The photometry time series; the blue
    shaded regions depict $\pm 1\sigma$ about the mean GP regression model of the binned
    photometry. \emph{Right column}: the generalized Lomb-Scargle periodogram of each season
    highlights the prevalence of photometric variations close to the measured rotation
    period and/or its first and second harmonics (vertical dashed lines).}
    \label{fig:WASPphot}
\end{figure*}

\subsection{Photometric rotation period}
\label{subsec:rotation}

\begin{figure}
    \centering
    \includegraphics[width=0.98\hsize]{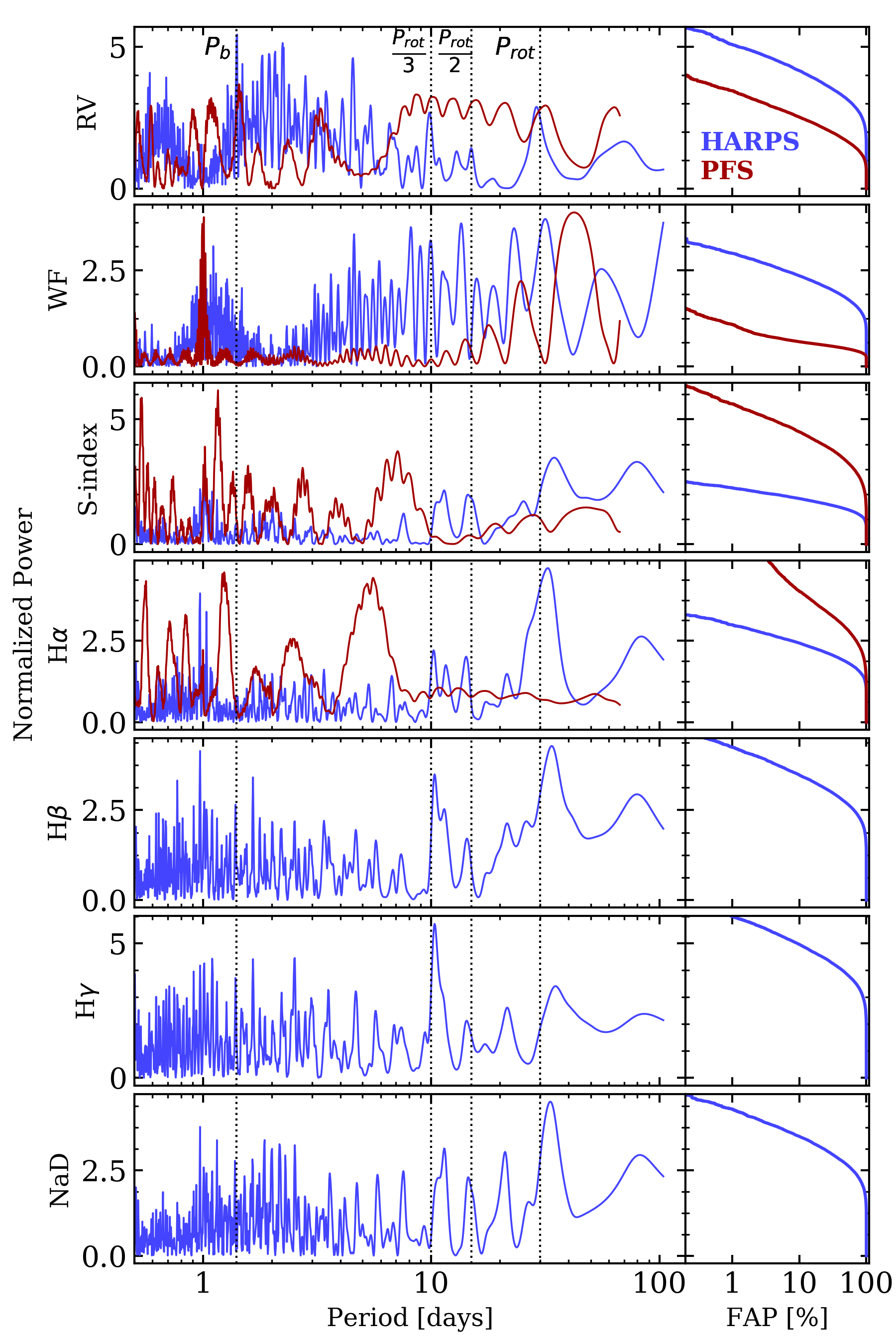}
    \caption{\emph{Left column}: Generalized Lomb-Scargle periodograms of 
    the HARPS and PFS RV time series, window functions, and the S-index, 
    $H\alpha$, $H\beta$, $H\gamma$, and sodium doublet activity indicators. 
    The vertical dotted lines highlight the locations of the L~168-9~b orbital 
    period, the stellar rotation period, and its first two harmonics. 
    \emph{Right column}: the false alarm probabilities computed 
    from bootstrapping with replacement.}
    \label{fig:GLSP}
\end{figure}

Knowledge of the stellar rotation period helps to disentangle
spurious RV signals arising from rotation and true RV signals due to orbital motion
\citep[e.g.,][]{Queloz2001,Cloutier2017b}. L~168-9 was photometrically monitored by WASP 
between May 2010 and December 2014 within five observing seasons each lasting approximately 
200 days in duration. The photometric precision, observing cadence, and baselines within 
the WASP fields are sufficient to detect quasi-periodic (QP) photometric variations of
L~168-9 due to active regions on the stellar surface rotating in and out of view at the
stellar rotation period \prot{.} The WASP photometry of L~168-9 is shown in 
Fig.~\ref{fig:WASPphot} along with the generalized Lomb-Scargle periodogram 
\citep[GLSP;][]{Zechmeister2009} of the photometry in each of the five WASP observing 
seasons. It is clear from the GLSPs that a strong periodicity exists within the data 
whose timescale is often $\sim 30$ days except for within the second WASP season wherein 
the dominant periodicity appears at the first harmonic of \prot{} $\sim 15$ days.

Given periodicities significantly detected in the WASP photometry, we proceeded to measure 
the photometric rotation period of L~168-9 \prot{} with each WASP camera and in each 
WASP field\footnote{At times, L~168-9 appeared within the fields-of-view of multiple 
WASP cameras.} in which L~168-9 was observed. As the photometric variations appear to 
vary nearly periodically, we modeled the photometry with a Gaussian process (GP) regression 
model and adopted a QP covariance kernel (see Eq.~\ref{eq:covariance})
\citep{Angus2018}. 
The covariance function's periodic timescale was a free parameter \prot{} for which the
posterior probability density function (PDF) was sampled using a Markov Chain 
Monte Carlo (MCMC) method (see Appendix~\ref{sec:GP}). We modeled each binned WASP light 
curve with a GP. We adopted a bin size of 1 day to reduce the computation time. 
In preliminary analyses, we also tested bin sizes of 0.25, 0.5, and 2 days 
and found that, probably due to the very large number of points, the recovered values of 
\prot{} were not very sensitive to this choice. 

After sampling the posterior PDFs of the GP hyperparameters, we arrived at point estimates 
of each parameter value based on the maximum a-posteriori values and 68 percent confidence
intervals. The resulting mean GP model of the data from each WASP observing season is
depicted in Fig.~\ref{fig:WASPphot} along with the corresponding $1\sigma$ confidence 
interval. Over the five observing seasons, the measured (median) rotation period 
of L~168-9 was \prot{} $=29.8\pm 1.3$ days. 

In principle, the WASP signal could have arisen from any star within the 48" extraction
aperture. However, L~168-9 is by far the brightest star in the aperture. Another concern
with any photometric signal with a period near 30 days is whether it was affected by
moonlight. To check on this possibility, we searched for modulations in the WASP data of
several stars of similar brightness within the surrounding 10 arcmin field, but did not find
any 30-d signals similar to the one that was seen for L~168-9. In any case, the star
location is far from the ecliptic, and moonlight contamination is not expected.
Furthermore, the modulation was sometimes seen at the 15-d first harmonic, which would not
be expected for moonlight. We can therefore be confident that the 30-d periodicity in the
WASP data arises from L~168-9. The $R^\prime_{HK}$ from HARPS spectra supports the 
obtained photometric rotation period, as $log(R^\prime_{HK})=-4.562\pm0.043$ (active star)
translates into \prot{} $=22\pm 2$ days using the $R^\prime_{HK}$vs.\prot{} relationship
from \cite{Astudillo2017a}.

\begin{figure*}[t]
\centering
\includegraphics[scale=0.5]{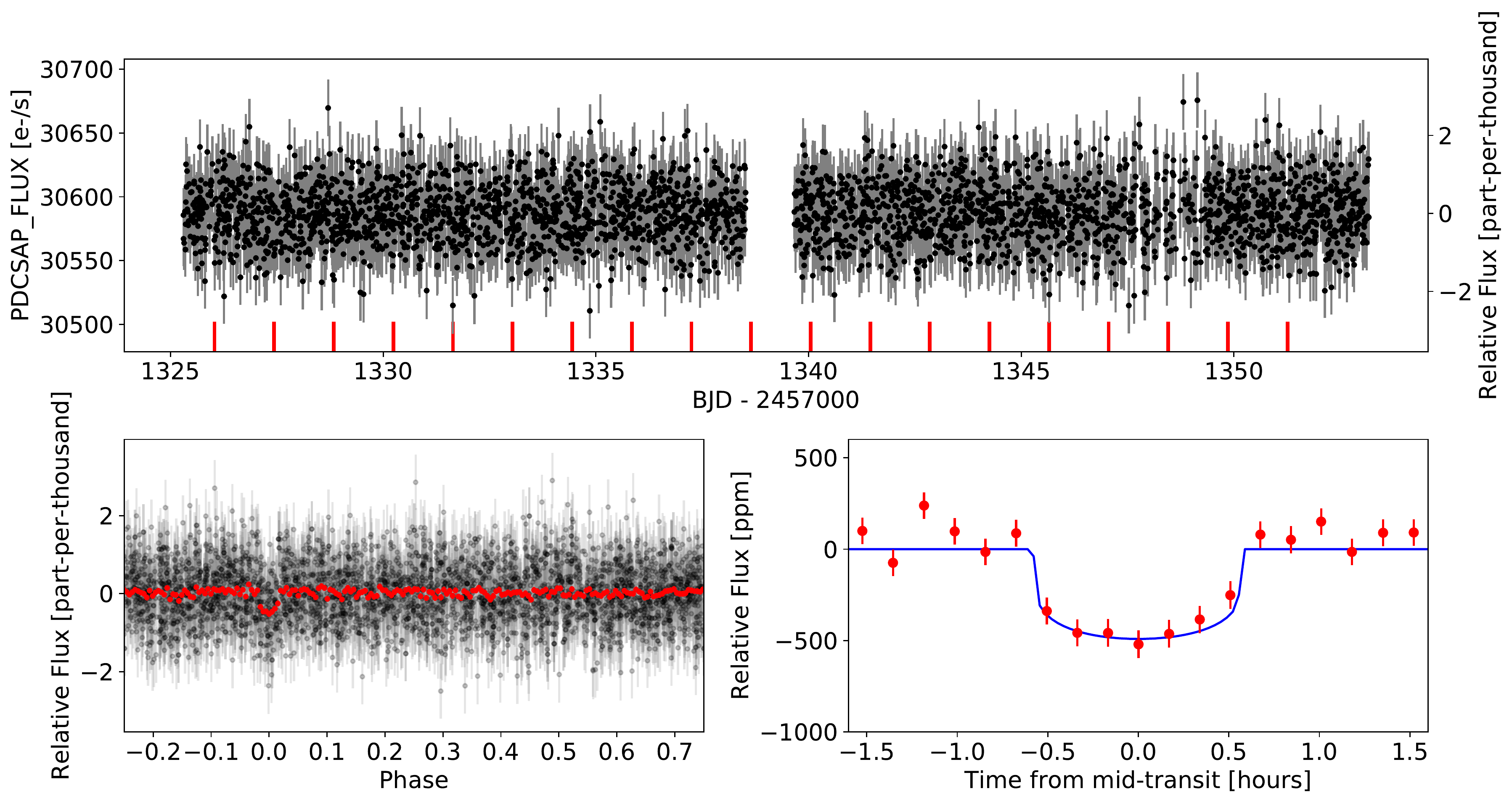}
\caption{The time series of TESS data for L~168-9. One fifth of original data are 
    plotted for visualization purposes. The de-trended TESS light curve is 
    shown in the \textit{upper panel}. Red vertical bars represent the transits 
    of the planet candidate. \textit{Bottom panel:} the phase folded normalized photometry.
    Red points correspond to binned data for illustrative purposes. The whole orbital phase 
    is shown to the left, while the right presents the transit phase, as well
    as the best transit model whose parameters come from the Table~\ref{tab:results}.}
\label{fig:tess_phot}
\end{figure*}

\subsection{Radial Velocity Periodogram Analysis}
\label{subsec:kep}

A first identification of significant periodicities in the HARPS and PFS RV time series is 
required in order to develop an accurate model of the observed RV variations. In a manner
similar to our analysis of the WASP photometry, we computed the GLSP of the following 
HARPS and PFS spectroscopic time series: the RVs, the window functions (WF), and the
S-index, $H\alpha$, $H\beta$, $H\gamma$, and the sodium doublet NaD activity indicators.
\cite{Astudillo-Defru2017b} details how these spectroscopic activity indicators were
derived. Each GLSP is shown in Fig.~\ref{fig:GLSP} along with a false alarm probability
(FAP) that was computed via bootstrapping with replacement using
$10^4$ iterations and normalizing each periodogram by its standard deviation.

Each of the GLSPs of the HARPS and PFS RV time series is dominated 
by noise and aliases arising from the respective WF. For example, the HARPS WF contains 
a forest of peaks with comparable FAP from $\sim 8$ days and extending out 
towards long periodicities. Similarly the GLSP of the PFS WF reveals a series 
of broad peaks for periodicities $\gtrsim 1$ day. These features, particularly 
those from the PFS WF, have clear manifestations in the GLSPs of their respective 
RV and activity indicator time series, thereby complicating the robust identification of 
periodocities in the data. However, strong peaks at the orbital period 
of L~168-9~b ($\sim 1.4$ days) are discernible in both RV time series at 
FAP $\sim 0.4\%$ and $\sim 0.3\%$ with HARPS and PFS respectively. This periodicity 
is not apparent in any of the ancillary activity indicators time series as expected 
for a signal originating from an orbiting planet. 

In addition to the signal from L~168-9~b, the HARPS RVs exhibit some power close 
to \prot{} and its second harmonic \prot{}$/3$. Although the FAPs of these 
periodicities over the full frequency domain are large, they each appear 
locally as strong periodicities as most of the power in the HARPS RVs exists
at $\lesssim 5$ days. The GLSP of the PFS RVs is much more difficult to interpret
at periodicities in the vicinity of \prot{} and its first and second harmonics 
due to strong aliases from the PFS WF. Due to these effects it is difficult to 
discern from the available PFS activity indices (i.e. S-index and $H\alpha$) 
whether or not a coherent activity signal is seen with PFS. Although each of 
the HARPS and PFS RV time series are significantly affected by sampling aliases, 
we do see evidence for L~168-9~b and rotationally modulated stellar activity 
in the RVs we endeavor to mitigate with our adopted model discussed 
in Sect.~\ref{subsec:model}.

\subsection{Radial velocity + transit model}
\label{subsec:model}

Guided by the periodicities in the HARPS and PFS RV time series, we proceeded 
to fit a model to the RVs including the effects of both stellar activity and the planet.
Following numerous successful applications on both Sun-like 
\citep[e.g.,][]{Haywood2014,Grunblatt2015,Faria2016,Lopezmorales2016,Mortier2016} 
and M dwarf stars \citep[e.g.,][]{Astudillo-Defru2017c,Bonfils2018,Cloutier2019a,Ment2019}, 
we adopted a QP kernel for the GP as a non-parametric model of the physical processes 
resulting in stellar activity. When used to model RV stellar activity, the QP 
covariance kernel is often interpreted as modelling the rotational component of 
stellar activity from active regions on the rotating stellar surface whose lifetimes 
typically exceed many rotation cycles on M dwarfs \citep{Giles2017} plus the 
evolutionary time scale of the active regions. The corresponding GP hyperparameters 
are described in detail in Appendix~\ref{sec:GP} and include each spectrograph's 
covariance amplitudes $a_{\text{HARPS}}$, $a_{\text{PFS}}$, the common exponential 
timescale $\lambda_{\text{RV}}$, the common coherence parameter $\Gamma_{\text{RV}}$, 
and the common periodic timescale $P_{\text{RV}}$ equal to the stellar rotation period \prot{.}

The planetary component attributed to the transiting planet L~168-9~b was fitted  
to the de-trended light curve with a \cite{MandelAgol2002} planetary transit model. 
The de-trended TESS light curve was produced by adjusting a QP GP systematic 
model to the photometry alone and with all the transits previously removed. 
The best QP GP model was subtracted to the entire TESS data set. The planetary component is 
modelled by a Keplerian solution parameterized by the planet's orbital period $P_b$, 
time of mid-transit $T_0$, RV semi-amplitude $K$, and the orbital parameters 
$h=\sqrt{e}\cos{\omega}$ and $k=\sqrt{e}\sin{\omega}$ where $e$ and $\omega$ 
are the planet's orbital eccentricity and argument of periastron respectively. 
In addition, our RV model contains each spectrograph's zero point velocity 
$\gamma_{\text{HARPS}}$, $\gamma_{\text{PFS}}$ and an additive scalar jitter 
$s_{\text{HARPS}}$, $s_{\text{PFS}}$ is account for any residual jitter that, 
unlike the stellar activity signal, is not temporally correlated. The complete 
RV model therefore contains fourteen model parameters.

To ensure self-consistent planet solutions between the available TESS transit 
data and the RV observations, we simultaneously fitted the de-trended light curve 
and the RVs. The common planetary parameters between these two data sets 
are $P_b$, $T_0$, $h$, and $k$.
The additional model parameters required to model the TESS transit light curve 
included an additive scalar jitter $s_{\text{TESS}}$, 
the baseline flux $\gamma_{\text{TESS}}$, the scaled semi-major axis $a/R_s$, 
the planet-star radius ratio $r_p/R_s$, the orbital inclination $i$, and the 
nearly-uncorrelated parameters $q_1$ and $q_2$ which are related to the 
quadratic limb darkening coefficients $u_1$ and $u_2$ via
\begin{align}
    q_1 &= (u_1+u_2)^2 \\
    q_2 &= \frac{u_1}{2(u_1+u_2)}.
    \label{eq:LDCs}
\end{align}

\noindent \citep{Kipping2013}. Thus we required eleven model parameters to describe the TESS 
transit light curve and a total of twenty-one model parameters of the joint RV + light curve 
data set: $\boldsymbol{\Theta}=\{a_{\text{HARPS}}, a_{\text{PFS}}, \lambda_{\text{RV}}, 
\Gamma_{\text{RV}}, P_{\text{RV}}, s_{\text{TESS}}, s_{\text{HARPS}}, s_{\text{PFS}}, \gamma_{\text{TESS}}, \gamma_{\text{HARPS}}$, $\gamma_{\text{PFS}}, 
P_b, T_0, K, h, k, a/R_s, r_p/R_s, i, q_1, q_2 \}$.

We sampled the posterior PDF of this 21-dimensional parameter space using an MCMC sampler. 
Details on the sampler and the adopted prior distributions on each model parameter are 
given in Appendix~\ref{sec:GP} and Table~\ref{tab:priors}.

\section{Results}
\label{sec:results}

\begin{figure}[t]
    \centering
    \includegraphics[width=1.0\hsize]{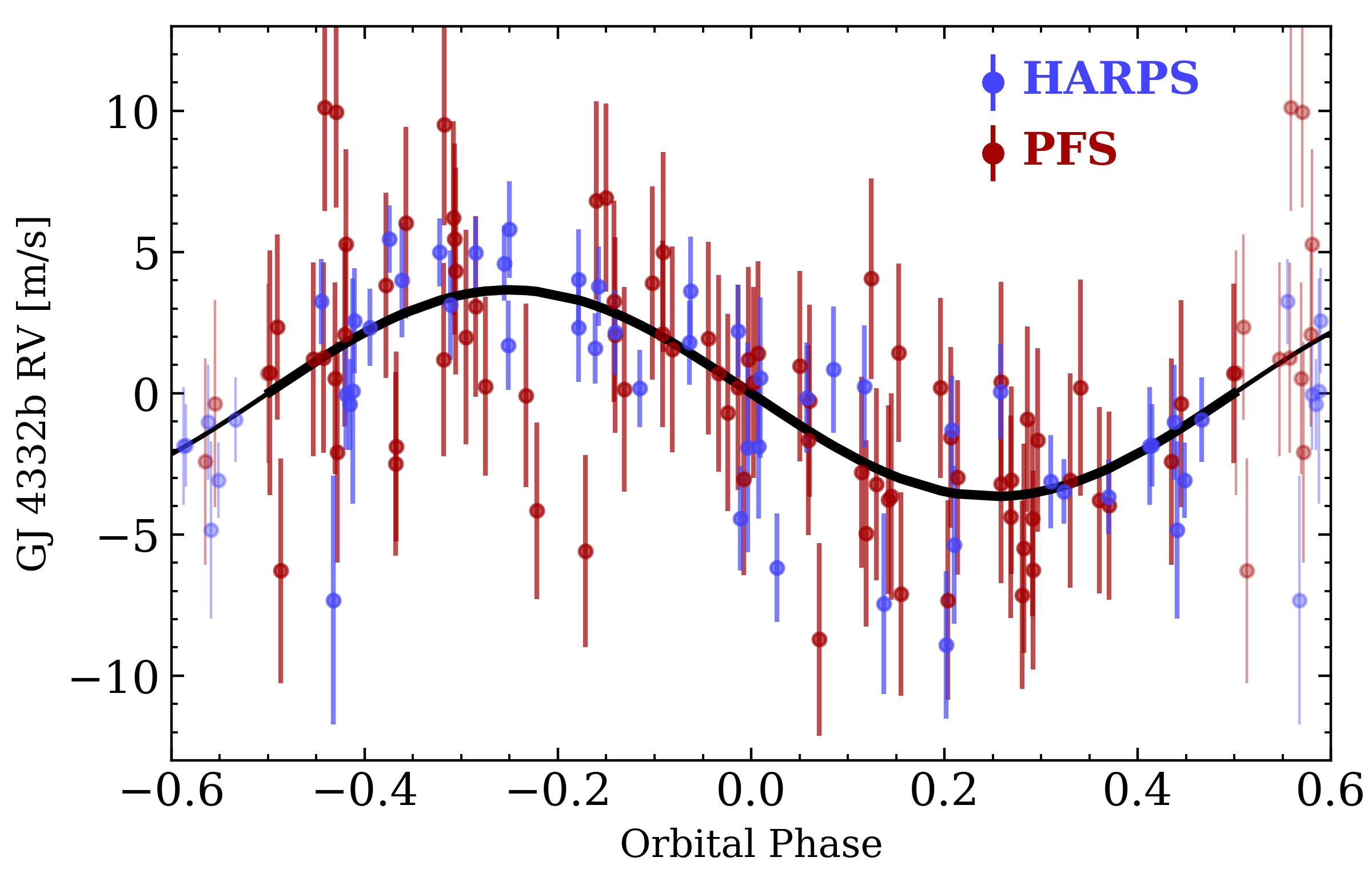}
    \caption{Phase folded radial velocity acquired HARPS (blue points) and PFS (red points)
    where the best GP model was subtracted. The black curve represents the maximum a-posteriori
    model adjusted to the data set.}
    \label{fig:phasedRV}
\end{figure}

In our analysis of the light curve and radial velocity time series of L~168-9, we evaluated 
the model presented in Sect.~\ref{subsec:model} on the separate RV data sets obtained with 
HARPS and PFS as well as the combined time series. Subtracting the model to the 46 HARPS 
radial velocity points reduces the dispersion to 3.37 m/s (equivalent to $\chi^2_{red}=3.5$),
while the dispersion of the 76 PFS residual points gives 4.05 m/s (translating into
$\chi^2_{red}=1.5$). The dispersion obtained from the 122 HARPS+PFS residual points is 
3.80 m/s ($\chi^2_{red}=2.2$).

From point estimates of the model parameters 
$\boldsymbol{\Theta}$ from our joint RV plus transit analysis with each of these 
input data sets we retrieved that L~168-9~b has a radius of 1.39$\pm$0.09 R$_\oplus$ and 
a mass of 4.60$\pm$0.56 M$_\oplus$, translating into a bulk mean density of 
$9.6^{+2.4}_{-1.8}$ g cm$^{-3}$. The planet is orbiting at 0.02091$\pm$0.00024 AU from 
the parent star, therefore the hot terrestrial planet has an equilibrium temperature between 
668 K and 965 K assuming a Venus-like and zero bond albedo, respectively.
Results for the entire set of parameters are reported in Table~\ref{tab:results}. 
Explicitly, we report the maximum a-posteriori value of each parameter along with its
16$^{th}$ and 84$^{th}$ percentiles, corresponding to a $1 \sigma$ confidence interval. 
We check for consistency of our joint analysis by performing the analysis for each instrument
independently. Table~\ref{tab:results} details the results from this test. We note that HARPS
and PFS results are in agreement within their uncertainties, translating into a robust detection
of the planetary signal in radial velocity data.

Figure~\ref{fig:tess_phot} show the TESS photometry and adjusted transit model and
Figure~\ref{fig:phasedRV} the phase folded radial velocity with the model that best 
fits the data.

\begin{table*}[t]
  \caption{Measured transit and RV model parameters of the L~168-9 planetary system.}
  \label{tab:results}
  \centering
  \begin{tabular}{lccc}
    \hline\noalign{\smallskip}
    Measured transit model parameters &  & TESS &  \\
    \noalign{\smallskip}
    \hline\noalign{\medskip}
    Baseline flux, $\gamma_{\text{TESS}}$ & \multicolumn{3}{c}{$0.00001\pm0.00029$}  \\
    \noalign{\smallskip}
    Orbital period, $P_b$ [days] & \multicolumn{3}{c}{$1.40150\pm 0.00018$}   \\
    \noalign{\smallskip}
    Time of mid-transit, $T_0$ [BJD-2,457,000] & \multicolumn{3}{c}{$1340.04781^{+0.00088}_{-0.00122}$}  \\
    \noalign{\smallskip}
    Scaled semi-major axis, $a/R_s$ & \multicolumn{3}{c}{$7.61\pm 0.31$}   \\
    \noalign{\smallskip}
    Planet-star radius ratio, $r_p/R_s$ & \multicolumn{3}{c}{$0.0212\pm 0.001$}   \\
    \noalign{\smallskip}
    Orbital inclination, $i$ [deg] & \multicolumn{3}{c}{$85.5^{+0.8}_{-0.7}$}  \\
    \noalign{\smallskip}
    Linear limb darkening coefficient, $q_1$ & \multicolumn{3}{c}{$0.397^{+0.125}_{-0.111}$}   \\
    \noalign{\smallskip}
    Quadratic limb darkening coefficient, $q_2$ & \multicolumn{3}{c}{$0.189^{+0.058}_{-0.056}$}   \\
    TESS additive jitter, $s_{\text{TESS}}$ & \multicolumn{3}{c}{$0.00003^{+0.00011}_{-0.00003}$}  \\
    \noalign{\smallskip}    
    \noalign{\medskip}    
    \hline\noalign{\smallskip}
    Radial velocity GP hyperparameters & HARPS+PFS & HARPS & PFS \\
    \noalign{\smallskip}
    \hline\noalign{\medskip}
    ln HARPS covariance amplitude,  $\ln{(a_{\text{HARPS}}/\text{m s}^{-1})}$ & $3.27^{+1.38}_{-1.03}$ & $3.24^{+1.24}_{-1.19}$ & - \\
    \noalign{\smallskip}
    ln PFS covariance amplitude, $\ln{(a_{\text{PFS}}/\text{m s}^{-1})}$ & $3.63^{+1.24}_{-1.08}$ & - & $4.01^{+1.23}_{-0.97}$ \\
    \noalign{\smallskip} 
    ln RV exponential timescale,
    $\ln{(\lambda_{\text{RV}}/\text{day})}$ & $11.90^{+2.81}_{-1.99}$ & $12.25^{+3.16}_{-2.18}$ & $13.3^{+2.02}_{-1.44}$\\
     \noalign{\smallskip} 
    ln RV coherence, $\ln{(\Gamma_{\text{RV}})}$ & $-0.09^{+0.20}_{-0.24}$ & $-0.42^{+0.41}_{-0.55}$ & $-0.22^{+0.39}_{-0.44}$ \\
     \noalign{\smallskip}
    ln RV periodic timescale,
    $\ln{(P_{\text{RV}}/\text{day})}$ & $3.47^{+0.02}_{-0.03}$ & $3.47^{+0.03}_{-0.03}$ & $3.48^{+0.04}_{-0.03}$ \\
    \noalign{\smallskip} 
    HARPS additive jitter, $s_{\text{HARPS}}$ [\mps{]} & $0.10^{+0.20}_{-0.09}$ & $0.86^{+0.98}_{-0.86}$ & - \\
    \noalign{\smallskip} 
    PFS additive jitter, $s_{\text{PFS}}$ [\mps{]} & $2.82\pm 0.42$ & - & $2.78\pm 0.30$  \\
    \noalign{\medskip}
    \hline\noalign{\smallskip}
    Measured RV model parameters & HARPS+PFS & HARPS & PFS \\
    \noalign{\smallskip}
    \hline\noalign{\medskip}
    HARPS zero point velocity, $\gamma_{\text{HARPS}}$ [km s$^{-1}$] & $29.7687\pm 0.0013$ & $29.7692\pm 0.0013$ & -  \\
    \noalign{\smallskip}
    PFS zero point velocity, $\gamma_{\text{PFS}}$ [km s$^{-1}$] & $0.00077 \pm 0.00186$ & - & $0.00063\pm 0.00142$  \\
    \noalign{\smallskip}
    Semi-amplitude, $K$ [\mps{]} & $3.66^{+0.47}_{-0.46}$ & $3.74^{+0.54}_{-0.59}$ & $3.26^{+0.52}_{-0.70}$  \\
    \noalign{\smallskip}
    $h=\sqrt{e}\cos{\omega}$ & $-0.10^{+0.17}_{-0.12}$ & $0.04^{+0.15}_{-0.18}$ & $-0.01^{+0.18}_{-0.21}$  \\
    \noalign{\smallskip}
    $k=\sqrt{e}\sin{\omega}$ & $0.00^{+0.17}_{-0.16}$ & $-0.09^{+0.22}_{-0.23}$ & $0.01^{+0.23}_{-0.30}$  \\
    \noalign{\medskip}
    \hline\noalign{\smallskip}
    Derived L~168-9~b parameters & HARPS+PFS+TESS & HARPS+TESS & PFS+TESS \\
    \noalign{\smallskip}
    \hline\noalign{\medskip}
    Semi-major axis, $a$ [AU] & $0.02091\pm 0.00024$ && \\
    Equilibrium temperature, $T_{\text{eq}}$ [K] &&& \\
    \hspace{10pt} Zero bond albedo & $965\pm 20$ && \\
    \hspace{10pt} Venus-like bond albedo = 0.77 & $668\pm 14$ && \\
    Planetary radius, $R_p$ [R$_{\oplus}$] & $1.39\pm 0.09$ && \\
    Planetary mass, $M_p$ [M$_{\oplus}$] & $4.60\pm 0.56$ & $4.74^{+0.71}_{-0.75}$ & $4.08^{+0.70}_{-0.90}$ \\
    Planetary bulk density, $\rho_p$ [g cm$^{-3}$] & $9.6^{+2.4}_{-1.8}$ & $10.0^{+2.5}_{-1.9}$ & $8.7^{+2.3}_{-2.1}$ \\
    Planetary surface gravity, $g_p$ [m s$^{-2}$] & $23.9^{+4.5}_{-3.9}$ & $24.4^{+4.8}_{-4.2}$ & $21.4^{+4.4}_{-4.8}$ \\
    Planetary escape velocity, $v_{\text{esc}}$ [km s$^{-1}$] & $20.5^{+1.4}_{-1.4}$ & $20.8^{+1.6}_{-1.7}$ & $19.4^{+1.6}_{-2.2}$ \\
    Orbital eccentricity, $e^\dagger$ & $< 0.21$ & $< 0.25$ & $< 0.26$ \\
    \noalign{\smallskip}\hline
  \end{tabular}
  
  \begin{list}{}{}
  \item[$^\dagger$] 95\% confidence interval.
  \end{list}
\end{table*}

\section{Discussion \& Conclusions}
\label{sec:conclusion}

L~168-9~b adds to the family of small ($<2R_\oplus$) transiting planets around bright 
($J<$ 8 mag) stars with mass measurements and contributes to the completion of the 
TESS Level One Science Requirement to detect and measure the masses of 50 small planets.
In particular, L~168-9~b is one of fourteen\footnote{L~98-59~bc \citep{Cloutier2019b}, GJ~357~b \citep{Luque2019}, HD~15337~b \citep{Dumusque2019}, HD~213885~b \citep{Espinoza2019}, GJ~9827~b \citep{Rice2019}, K2-265~b \citep{Lam2018}, K2-141~b \citep{Barragan2018}, K2-229~b \citep{Santerne2018}, HD~3167~b \citep{Gandolfi2017}, K2-106~b \citep{Guenther2017}, TRAPPIST-1~fh \citep{Wang2017}, HD~219134~b \citep{Motalebi2015}} 
likely rocky planets without primordial hydrogen-helium envelopes 
-- that is,  with a radius $<1.8R_\oplus$ -- 
for which the mass has been measured with an uncertainty smaller than 33\%.
Thus, our result represents progress toward the understanding of the 
transition between super-Earths and mini-Neptunes previously reported in the radii of
planets \citep[e.g.][]{Fulton2017,Cloutier2019c} but, here, including the information on mass.

\begin{figure}[t]
    \centering
    \includegraphics[scale=0.6]{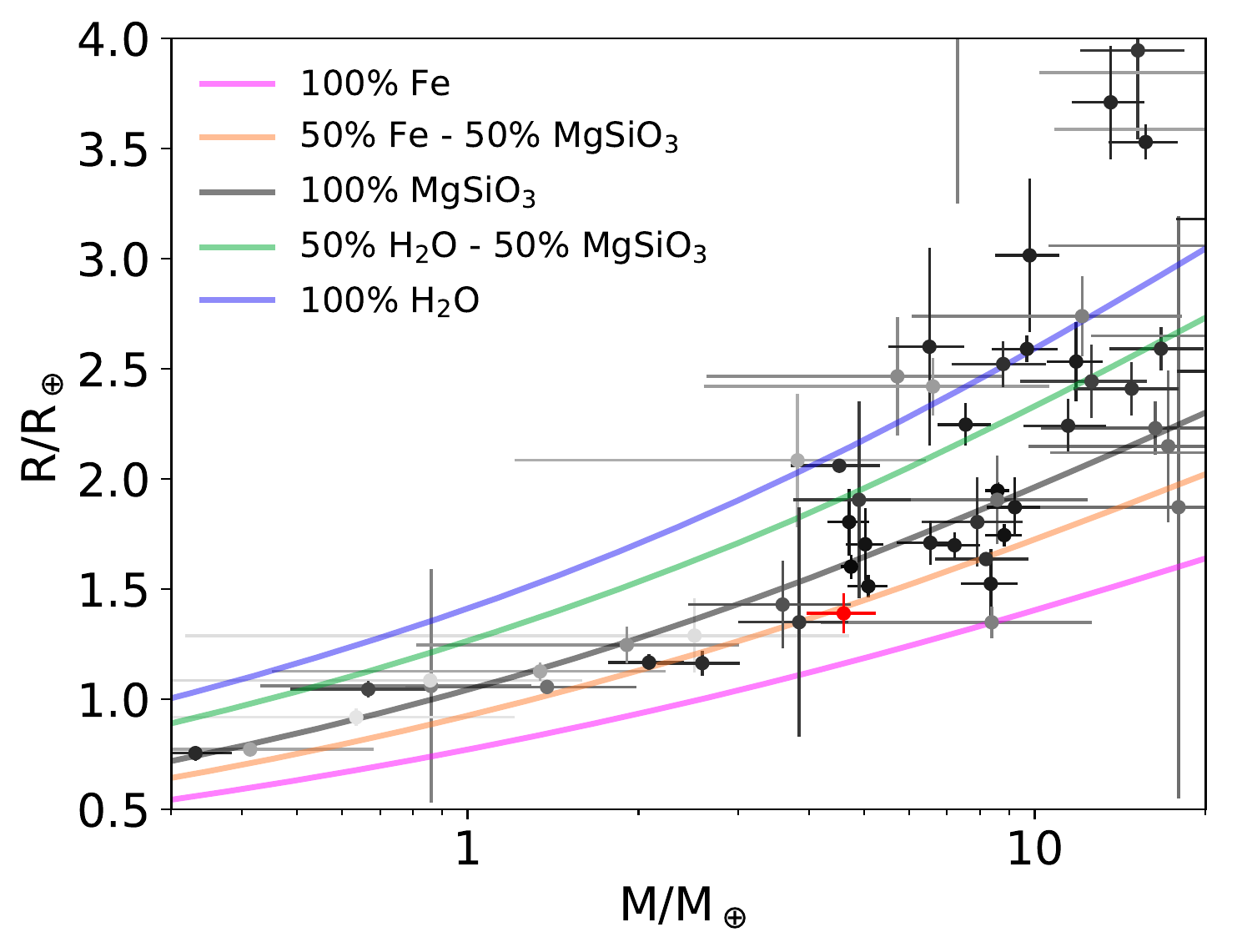}
    \caption{The mass-radius diagram showing L~168-9~b (red circle) in the context of known
    exoplanets. The transparency of each point is proportional to its associated mass
    uncertainty. Error bars correspond to $1 \sigma$ uncertainties. Different models for 
    the bulk composition are plotted where the legend details the fraction of iron,
    silicates, and/or water for each color-coded curve.}
    \label{fig:MR}
\end{figure}

Figure~\ref{fig:MR} shows the mass~-~radius diagram centered in the sub-Earth to
mini-Neptune regime. With about twice the Earth average density, L~168-9~b bulk density 
is compatible with a terrestrial planet with an iron core (50\%) surrounded by a mantle of
silicates (50\%). In this diagram the detected planet is located in an interesting place:
for masses lower than that of L~168-9~b the great majority of planets are consistent with a 
50\% Fe--50\% MgSiO3 or 100\% MgSiO3 bulk composition, while for higher planetary masses
there is a great diversity of density. Being one of the densest planets for masses greater
than 4M$_\oplus$, L~168-9~b can help to define the mass limits of the rocky planets
population.

Good targets for atmospheric characterization with transmission/emission spectroscopy are 
those transiting nearby, bright stars ($J<$ 8 mag). There are currently 11 small planets
detected transiting a bright star, according to the
NASA Exoplanet Archive\footnote{\url{https://exoplanetarchive.ipac.caltech.edu/}}, two of
them orbit M~dwarfs. Overall, there would not be a large number of small, transiting
planets around nearby, bright M~dwarfs. There are roughly 200 M~dwarfs within the 25~pc
solar neighborhood with $J<$ 8 mag, about 2/3 of which are single stars
\citep{Winters:2015,Winters:2019}. Considering the occurrence rate of small planets with
orbital period smaller than 10 days from \cite{Dressing:2015} and combining with the transit
probability of such planets (about a couple \% to $\sim$20\%), there would be a few to up to
about twenty such planets.

The measured properties of L~168-9~b and its host star make it a promising target for 
the atmospheric characterization of a terrestrial planet via either transmission 
or emission spectroscopy measurements with JWST \citep{Morley2017,Kempton2018} and/or
thermal phase curve analysis to infer the absence of a thick atmosphere
\citep[e.g.,][]{Seager2009}. 
Its transmission and emission spectroscopy metrics from \cite{Kempton2018} 
are reported in Table~\ref{tab:atmosphere} and compared to other confirmed transiting 
terrestrial planets with known masses that are of interest for atmospheric characterization.
Based on this assessment, L~168-9~b is an excellent candidate for emission spectroscopy or
for detecting the planetary day-side phase curve as recently done for the similar planet,
LHS~3844~b \citep{Kreidberg2019}.

\begin{table*}[t]
  \caption{Prospects of atmospheric characterization of confirmed terrestrial planets including L~168-9~b. 
  TSM and ESM correspond to transmission and emission spectroscopy metrics, respectively.}
  \label{tab:atmosphere}
  \centering
  \small
  \begin{tabular}{lcccccccccccccc}  
    \hline\noalign{\smallskip}
    Star ID  & R$_{p}$ & M$_{p}$ & P & a& T$_{eff}$ & T$_{eq}$ & T$_{day}$ & J & K$_{\rm s}$  & R$_\star$ & M$_\star$ & TSM & ESM & Ref.\\
    units  & [R$_\oplus$] & [M$_\oplus$] & [days] & [AU]& [K] & [K] & [K] & [mag] & [mag]  & [R$_\odot$] & [M$_\odot$] &  &  & \\
    \noalign{\smallskip}\hline\noalign{\smallskip}
    L~168-9~b   & 1.39 & 4.39 & 1.401 & 0.021 & 3743 & 963.01 & 1059.31& 7.941 & 7.0819 & 0.60 & 0.62 & 8.025  & 9.692& \\
    LHS~3844~b   & 1.30 & --   & 0.463 & 0.006 & 3036 & 805.90 & 886.49 & 10.046& 9.145  & 0.19 & 0.15 &     -- & 29.004 & Kr2019\\
    GJ~1132~b    & 1.13 & 1.66 & 1.629 & 0.015 & 3270 & 590.61 & 649.67 & 9.245 & 8.322  & 0.21 & 0.18 & 31.166 & 9.872 & Bo2018\\
    L~98-59~c    & 1.35 & 2.17 & 3.690 & 0.032 & 3412 & 515.31 & 566.84 & 7.933 & 7.101  & 0.31 & 0.31 & 29.168 & 6.696 & Cl2019b\\
    LTT~1445A~b  & 1.38 & 2.20 & 5.359 & 0.038 & 3335 & 433.34 & 476.68 & 7.29  & 6.5    & 0.28 & 0.26 & 44.976 & 6.382 & Wi2019\\
    TRAPPIST-1~b & 1.09 & 1.02 & 1.511 & 0.011 & 2559 & 402.38 & 442.62 & 11.4  & 10.3   & 0.12 & 0.08 & 36.914 & 4.007 & Gi2017\\
    LHS~1140~c   & 1.28 & 1.81 & 3.778 & 0.027 & 3216 & 436.43 & 480.07 & 9.612 & 8.821  & 0.21 & 0.18 & 25.225 & 3.401 & Me2019\\
    \noalign{\smallskip}\hline
  \end{tabular}

  \begin{list}{}{}
  \item Reference notes: Kr2019~--~\citep{Kreidberg2019}; Bo2018~--~\citet{Bonfils2018}; Cl2019b~--~\citet{Cloutier2019b}; Wi2019~--~\citet{Winters2019}; Gi2017~--~\citet{Gillon2017}; Me2019~--~\citet{Ment2019}; 
  \end{list}
\end{table*}

\begin{acknowledgements}
  N. A.-D. acknowledges the support of FONDECYT project 3180063.
  J.K.T. acknowledges that support for this work was provided by NASA through Hubble
  Fellowship grant HST-HF2-51399.001 awarded by the Space Telescope Science Institute, 
  which is operated by the Association of Universities for Research in Astronomy, Inc., 
  for NASA, under contract NAS5-26555.
  R.B.\ acknowledges support from FONDECYT Post-doctoral Fellowship Project 3180246, and
  from the Millennium Institute of Astrophysics (MAS).
  X.B. and J.M-A. acknowledge funding from the European Research Council
  under the ERC Grant Agreement n. 337591-ExTrA.
  X.D.; X.B.; T.F.; et L.M. acknowledge the  support by the French National Research Agency
  in the  framework of the Investissements d’Avenir program (ANR-15-IDEX-02), through the
  funding of the "Origin of Life" project of the Univ. Grenoble-Alpes.”
  LM acknowedge the support of the Labex OSUG@2020 (Investissements d’avenir -- ANR10
  LABX56).
  JRM acknowledges CAPES, CNPq and FAPERN brazilian agencies.
  This work was supported by FCT/MCTES through national funds and by FEDER - 
  Fundo Europeu de Desenvolvimento Regional through COMPETE2020 - Programa Operacional
  Competitividade e Internacionalização by these grants: UID/FIS/04434/2019;
  PTDC/FIS-AST/32113/2017 \& POCI-01-0145-FEDER-032113; PTDC/FIS-AST/28953/2017 \&
  POCI-01-0145-FEDER-028953.
  T.H. acknowledges support from the European Research Council under the Horizon 2020
  Framework Program via the ERC Advanced Grant Origins 83 24 28.
  REM acknowledges support by the BASAL Centro de Astrof\'isica y Tecnolog\'ias Afines (CATA) and FONDECYT
  1190621.
  A.J.\ acknowledges support from FONDECYT project 1171208 and by the Ministry for the Economy, Development, and Tourism's Programa Iniciativa Cient\'{i}fica Milenio through grant IC\,120009, awarded to the Millennium Institute of Astrophysics (MAS).
  JGW is supported by a grant from the John Templeton Foundation. The opinions expressed here are
  those of the authors and do not necessarily reflect the views of the John Templeton Foundation.
  Work J.N.W.\ was partly funded by the Heising-Simons Foundation.
  The authors would like to acknowledge Zachary Hartman for his help conducting the 
  Gemini-South/DSSI observations. Some of the work here is based on observations 
  obtained at the Gemini Observatory, which is operated by the Association of Universities 
  for Research in Astronomy, Inc., under a cooperative agreement with the NSF on 
  behalf of the Gemini partnership: the National Science Foundation (United States), 
  National Research Council (Canada), CONICYT (Chile), Ministerio de Ciencia, 
  Tecnolog\'{i}a e Innovaci\'{o}n Productiva (Argentina), 
  Minist\'{e}rio da Ci\^{e}ncia, Tecnologia e Inova\c{c}\~{a}o (Brazil), and Korea 
  Astronomy and Space Science Institute (Republic of Korea). 
  This work makes use of observations from the LCOGT network.
  Funding for the TESS mission is provided by NASA's Science Mission directorate.
  We acknowledge the use of public TESS Alert data from pipelines at the TESS 
  Science Office and at the TESS Science Processing Operations Center. 
  This research has made use of the Exoplanet Follow-up Observation Program website, 
  which is operated by the California Institute of Technology, under contract with 
  the National Aeronautics and Space Administration under the Exoplanet Exploration Program.
  Resources supporting this work were provided by the NASA High-End Computing (HEC) Program
  through the NASA Advanced Supercomputing (NAS) Division at Ames Research Center for the
  production of the SPOC data products.
  This paper includes data collected by the TESS mission, which are publicly available 
  from the Mikulski Archive for Space Telescopes (MAST).
  This research has made use of the NASA Exoplanet Archive, which is operated by the
  California Institute of Technology, under contract with the National Aeronautics and 
  Space Administration under the Exoplanet Exploration Program.
\end{acknowledgements}

% WARNING
%-------------------------------------------------------------------
% Please note that we have included the references to the file aa.dem in
% order to compile it, but we ask you to:
%
% - use BibTeX with the regular commands:
\bibliographystyle{aa} % style aa.bst
\bibliography{TOI134} % your references Yourfile.bib
%
% - join the .bib files when you upload your source files
%-------------------------------------------------------------------

\begin{appendix} %First appendix

\section{Gaussian process model}
\label{sec:GP}

Gaussian process (GP) regression is widely used in the exoplanet community as a non-parametric
Bayesian approach to model temporally correlated stellar activity signals in RV data. 
These rotationally-modulated activity signals prohibit the accurate measurement of planetary parameters and often produce in planetary false positives.

Here we model the RV activity signals of L~168-9 as a stochastic process whose temporal
evolution is well-described by a quasi-periodic covariance kernel. A GP with a
quasi-periodic covariance kernel $k(t_i,t_j)$ is included in our joint model describing 
the data RV and transit data and takes the following form:

\begin{equation}
    \label{eq:covariance}
    k(t_i,t_j) = a^2 \exp{\left[ -\frac{(t_i-t_j)^2}{2\lambda^2} -\Gamma^2 \sin^2{\left( \frac{\pi |t_i-t_j|}{P_{\text{GP}}} \right)} \right]}
\end{equation}

\noindent and is described by the covariance amplitude $a$, the exponential
evolutionary timescale $\lambda$, the coherence $\Gamma$, and the periodic timescale $P_{\text{GP}}$.

As usual in the Bayesian context, prior probability density functions (PDF) for the
hyperparameters $\{ a,\ \lambda,\ \Gamma,\ P_{\text{GP}} \}$ are listed in
Table~\ref{tab:priors} for the multiple
GPs applied our data analysis. The posterior PDF are sampled through a 
Markov chain Monte Carlo
\citep[MCMC,][]{GoodmanAndWeare2010}; in particular we used the \texttt{emcee} ensemble
sampler \citep{Foremanmackey2013}. The sampling of the joint posterior is made
with the Gaussian ln likelihood function given by

\begin{equation}
    \ln{\mathcal{L}} = -\frac{1}{2} \left( y^{\text{T}}\cdot K \cdot y + \ln{\text{det}K} + N\ln{2\pi} \right)
\end{equation}

\noindent where $y$ is the vector of $N$ measurement taken at times
$t=\{t_1,t_2,\dots,t_N \}$ and the $N \times N$ covariance matrix $K$ is given by

\begin{equation}
    K_{ij} = k(t_i,t_j) + \delta_{ij}[\sigma(t_i)^2 + s^2].
\end{equation}

\noindent being $\delta_{ij}$ the Kronecker delta that adds the measurement uncertainties
$\sigma$ to the diagonal elements of $K$ and includes an additive jitter factor $s$.

\begin{table}[t]
  \caption{L~168-9 model parameter priors.}
  \label{tab:priors}
  \centering
  \small
  \begin{tabular}{lc}  
    \hline\noalign{\smallskip}
    Parameter & Prior \\
    \hline\noalign{\smallskip}
    \multicolumn{2}{c}{Photometry model} \\
    Baseline flux, $\gamma_{\text{TESS}}$  & $\mathcal{U}(-0.1,0.1)$  \\
    Limb darkening coefficient, $u_1$ &$\mathcal{U}(0.1,0.4)$ \\
    Limb darkening coefficient, $u_2$ & $\mathcal{U}(0.25,0.55)$  \\
    TESS additive jitter, $s_{\text{TESS}}$ & $\mathcal{J}(10^{-2},1)$   \\
    \noalign{\smallskip}
     &  \\
    \noalign{\smallskip}
    \multicolumn{2}{c}{RV model} \\
    \noalign{\smallskip}
    HARPS zero point velocity, $\gamma_{\text{0,HARPS}}$ [\mps{]} & $\mathcal{U}(-1,1)$ \\
    PFS zero point velocity, $\gamma_{\text{0,PFS}}$ [\mps{]} & $\mathcal{U}(-1,1)$ \\
    ln HARPS covariance amplitude, $\ln{(a_{\text{HARPS}}/\text{m s}^{-1})}$ & $\mathcal{U}(-5,10)$ \\
    ln PFS covariance amplitude, $\ln{(a_{\text{PFS}}/\text{m s}^{-1})}$ & $\mathcal{U}(-5,10)$ \\

    ln RV exponential time scale, $\ln{(\lambda_{\text{RV}}/\text{day})}$ & From H$\alpha$ training \\
    ln RV coherence, $\ln{(\Gamma_{\text{RV}})}$ & From H$\alpha$ training \\
    ln RV periodic timescale, $\ln{(P_{\text{RV}}/\text{day})}$ & From H$\alpha$ training \\
    HARPS additive jitter, $s_{\text{HARPS}}$ [\mps{]} & $\mathcal{J}(10^{-2},10)$ \\
    PFS additive jitter, $s_{\text{PFS}}$ [\mps{]} & $\mathcal{J}(10^{-2},10)$ \\
    \noalign{\smallskip}
    \multicolumn{2}{c}{\emph{L~168-9~b (TOI-134.01)}} \\
    \noalign{\smallskip}
    Orbital period, $P_b$ [days] & $\mathcal{U}(1.370,1.405)$ \\
    Time of mid-transit, $T_{0,b}$ [BJD - 2,457,000] & $\mathcal{U}(1338, 1342)$ \\
    Scaled semi-major axis, $a/R_s$ & $\mathcal{G}(7.542, 0.27)$ \\
    Planet-star radius ratio, $r_p/R_s$ & $\mathcal{U}(0,0.1)$ \\
    Orbital inclination, $i$ [deg] &  $\mathcal{U}(75,105)$ \\
    Semi-amplitude, $K_b$ [\mps{]} & $\mathcal{J}(10^{-2},10)$ \\
    $h_b=\sqrt{e_b}\cos{\omega_b}$ & $\mathcal{U}(-1,1)$ \\
    $k_b=\sqrt{e_b}\sin{\omega_b}$ & $\mathcal{U}(-1,1)$ \\
    \noalign{\smallskip}
    
    \hline\noalign{\smallskip}
  \end{tabular}
\end{table}

\section{Spectroscopic Data}

\begin{table*}
\small
\caption{HARPS radial velocity time series and spectroscopic activity indicators for L~168-9 (minimal, full version available at the CDS).  }
\label{tab:rvHARPS}
\begin{tabular*}{\hsize}{@{\extracolsep{\fill}}lllllllllllll}

\noalign{\smallskip}
\hline\hline
\noalign{\smallskip}

BJD & RV &  $\sigma_{RV}$ & H$\alpha$ & $\sigma_{H \alpha}$ & H$\beta$ & $\sigma_{H \beta}$ & H$\gamma$ & $\sigma_{H \gamma}$ & NaD & $\sigma_{NaD}$ & S & $\sigma_{S}$ \\
\text{[-2450000]} & $[ms^{-1}]$ & $[ms^{-1}]$ &&&&&&&&&& \\

\noalign{\smallskip}
\hline
\noalign{\smallskip}

4664.954608 & 29781.65 & 3.48 & 0.06224 & 0.00025 & 0.04679 & 0.00052 & 0.10087 & 0.00131 & 0.01232 & 0.00020 & 1.616 & 0.068\\
4991.927461 & 29786.04 & 2.95 & 0.06074 & 0.00021 & 0.04632 & 0.00044 & 0.10385 & 0.00114 & 0.01172 & 0.00016 & 2.167 & 0.073\\
8367.523380 & 29769.97 & 1.37 & 0.05962 & 0.00011 & 0.04606 & 0.00019 & 0.10582 & 0.00052 & 0.01166 & 0.00007 & 1.892 & 0.018\\
8367.551447 & 29773.09 & 1.19 & 0.05928 & 0.00010 & 0.04534 & 0.00016 & 0.10573 & 0.00045 & 0.01158 & 0.00006 & 1.826 & 0.014\\
8367.623921 & 29772.57 & 1.20 & 0.05944 & 0.00010 & 0.04562 & 0.00017 & 0.10639 & 0.00046 & 0.01164 & 0.00006 & 1.868 & 0.014\\

\noalign{\smallskip}
\hline

\end{tabular*}
\end{table*}

\begin{table*}
\small
\caption{PFS radial velocity time series and spectroscopic activity indicators (minimal, full version available at the CDS).}
\label{tab:rvPFS}
\begin{tabular*}{\hsize}{@{\extracolsep{\fill}}lllll}

\noalign{\smallskip}
\hline\hline
\noalign{\smallskip}

BJD & RV &  $\sigma_{RV}$ & H$\alpha$ & S  \\
\text{[-2450000]} & $[ms^{-1}]$ & $[ms^{-1}]$ && \\

\noalign{\smallskip}
\hline
\noalign{\smallskip}

8409.51799 & 8.80 & 1.74 & 1.293 & 0.05807  \\
8409.53206 & 4.10 & 1.74 & 1.334 & 0.05813  \\
8409.60491 & -3.11 & 1.71 & 1.285 & 0.05879 \\
8409.61960 & 4.80 & 1.75 & 1.340 & 0.05901  \\
8409.67495 & 8.23 & 2.09 & 1.311 & 0.05976  \\

\noalign{\smallskip}
\hline

\end{tabular*}
\end{table*}

\end{appendix}

\end{document}